%% file: main.tex
\documentclass[%
 reprint,
 prl,
 showpacs,
 amsmath,amssymb,
 aps,
 lengthcheck,
floatfix,
]{revtex4-1}

\usepackage{bibunits}

\usepackage{graphicx}
\usepackage{dcolumn}
\usepackage{bm}
\usepackage{hyperref}
\usepackage[mathlines]{lineno}

\usepackage{soul}

\usepackage{graphicx}
\usepackage{times}
\usepackage[normalem]{ulem}
\usepackage{color}
\usepackage{cellspace}
\newcommand{\comment}[1]{}
\renewcommand\sout{\bgroup \color{red} \ULdepth=-.5ex \ULset}
\def\simge{\mathrel{\rlap{\raise 0.511ex
     \hbox{$>$}}{\lower 0.511ex \hbox{$\sim$}}}}
\def\simle{\mathrel{\rlap{\raise 0.511ex
      \hbox{$<$}}{\lower 0.511ex \hbox{$\sim$}}}}

\graphicspath{{./}{figures/}}

\begin{document}

\title{Tidal Deformabilities and Radii of Neutron Stars from the Observation of GW170817}

\author{Soumi De$^{1}$}
\author{Daniel Finstad$^1$}
\author{James M. Lattimer$^2$}
\author{Duncan A. Brown$^1$}
\author{Edo Berger$^3$}
\author{Christopher M. Biwer$^{1,4}$}

\affiliation{$^1$ Department of Physics, Syracuse University, Syracuse, NY 13244, USA}
\affiliation{$^2$ Department of Physics and Astronomy, Stony Brook University, Stony Brook, NY 11794-3800, USA}
\affiliation{$^3$ Harvard-Smithsonian Center for Astrophysics, 60 Garden Street, Cambridge,
Massachusetts 02139, USA}
\affiliation{$^4$ Applied Computer Science (CCS-7), Los Alamos National Laboratory, Los Alamos, NM, 87545, USA}
 
\begin{abstract}
We use gravitational-wave observations of the binary neutron star merger GW170817 to explore the tidal deformabilities and radii of neutron stars.
We perform Bayesian parameter estimation with the source location and distance informed by electromagnetic observations. We also assume that the two stars have the same equation of state; we demonstrate that for stars with masses comparable to the component masses of GW170817, this is effectively implemented by assuming that the stars' dimensionless tidal deformabilities are determined by the binary's mass ratio $q$ by $\Lambda_1/\Lambda_2 = q^6$. We investigate different choices of prior on the component masses of the neutron stars. We find that the tidal deformability and 90\% credible interval is $\tilde{\Lambda}=222^{+420}_{-138}$ for a uniform component mass prior, $\tilde{\Lambda}=245^{+453}_{-151}$ for a component mass prior informed by radio observations of Galactic double neutron stars, and $\tilde{\Lambda}=233^{+448}_{-144}$ for a component mass prior informed by radio pulsars. We find a robust measurement of the common areal radius of the neutron stars across all mass priors of $8.9 \le \hat{R} \le 13.2$~km, with a mean value of $\langle \hat{R} \rangle = 10.8$~km. Our results are the first measurement of tidal deformability with a physical constraint on the star's equation of state and place the first lower bounds on the deformability and areal radii of neutron stars using gravitational waves.
\end{abstract}

\pacs{95.85.Sz, 26.60.Kp, 97.80.-d}
\maketitle

\begin{bibunit}[main]

\textit{Introduction.}---On August 17, 2017 LIGO and Virgo observed gravitational waves from a binary neutron star coalescence, GW170817~\cite{TheLIGOScientific:2017qsa}.
This observation can be used to explore the equation of state (EOS) of matter at super-nuclear densities~\cite{thorne.k:1987,Read:2009yp}. This information is encoded as a change in gravitational-wave phase evolution caused by the tidal deformation of the  neutron stars~\cite{Flanagan:2007ix}. 
At leading order, the tidal effects are imprinted in the gravitational-wave signal through the binary tidal deformability~\cite{Flanagan:2007ix,Hinderer:2007mb}
\begin{equation}
\tilde{\Lambda}
= \frac{16}{13}\frac{(12q+1)\Lambda_1+(12+q)q^4\Lambda_2}{(1+q)^5}, \label{eq:lambda_t0}
\end{equation}
where $q = m_2/m_1 \leq 1$ is the binary's mass ratio [cf. Eq.~(34) of Ref.~\cite{Gralla:2017djj}]. The deformability of each star is
\begin{equation}
\Lambda_{1,2}=\frac{2}{3}k_2\left(\frac{R_{1,2}c^2}{Gm_{1,2}}\right)^5, \label{eq:lambda_12}
\end{equation}
where $k_2$ is the tidal Love number~\cite{Flanagan:2007ix,Hinderer:2007mb}, which depends on the star's mass and the EOS. $R_{1,2}$ and $m_{1,2}$ are the areal radii and masses of the neutron stars, respectively. 

In the results of Ref.~\cite{TheLIGOScientific:2017qsa},
the priors on $\Lambda_{1,2}$ are taken to be completely uncorrelated,
which is equivalent to assuming that each star may have a different
EOS. Here, we reanalyze the gravitational-wave data using Bayesian
inference \cite{Biwer:2018osg,alex_nitz_2018_1208115,emcee} to measure the tidal deformability, using a
correlation between $\Lambda_1$ and $\Lambda_2$ which follows from the
assumption that both stars have the same EOS.  We repeat
our analysis without the common EOS constraint and calculate the 
Bayes factor that compares the evidences for these two models.
We also fix the sky position and
distance from electromagnetic observations~\cite{Soares-Santos:2017lru,Cantiello:2018ffy}.
We study the effect of
the prior for the component masses by performing analyses with three
different priors: the first is uniform between 1 and $2M_\odot$, the
second is informed by radio observations of double neutron star
binaries, and the third is informed by the masses of isolated pulsars~\cite{Ozel:2016oaf}.

\begin{figure}[t]
  \includegraphics[width=\columnwidth]{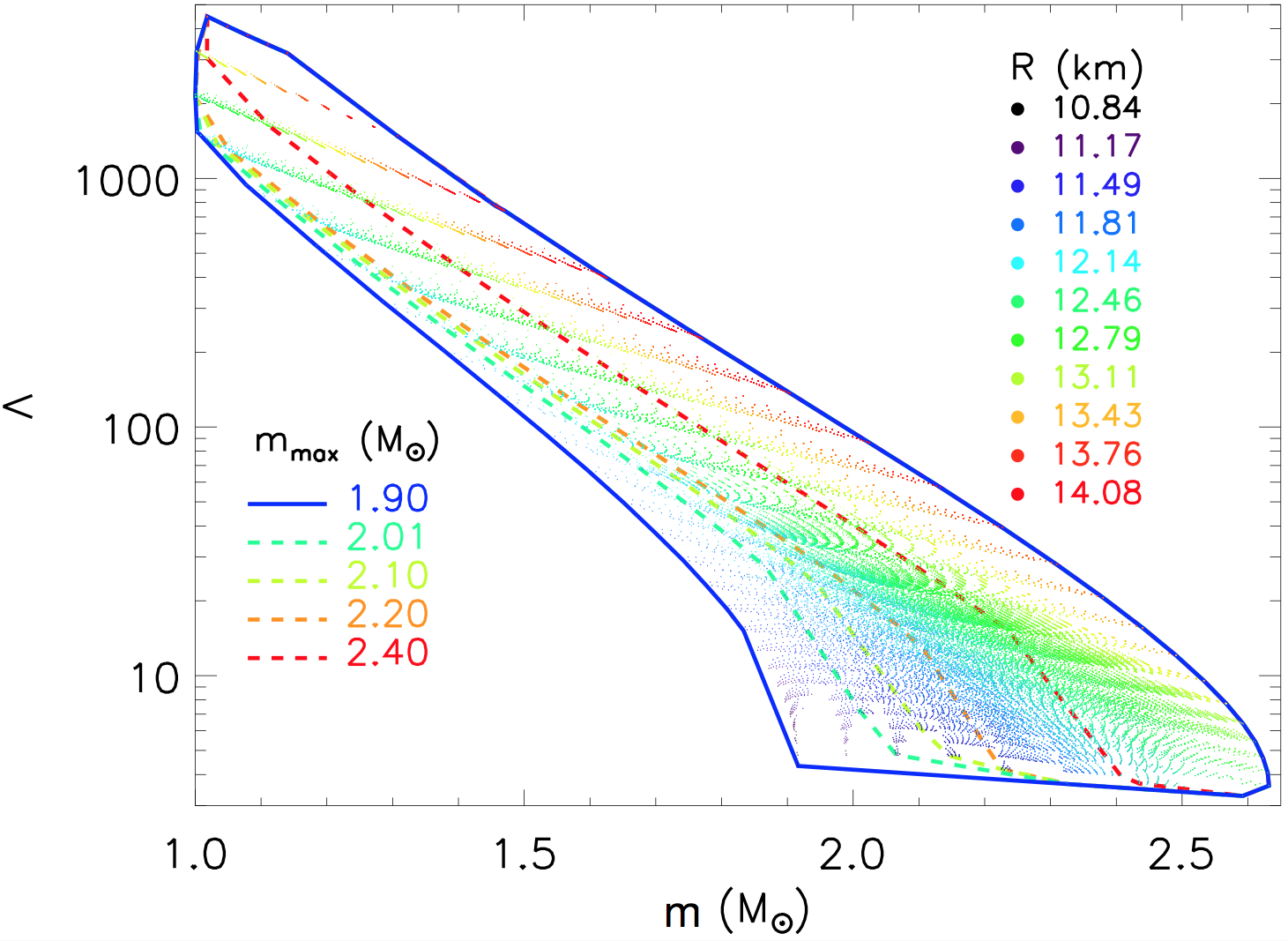}
  \caption{The tidal deformability $\Lambda$ as a function of mass for physically realistic polytropes. A TOV integration with each EOS parameter set results in a series of values of $\Lambda(m)$ that are shown as points colored by their radii $R$. Dashed curves are lower bounds to $\Lambda$ for a given mass $m$ which vary depending on the assumed lower limit to the neutron star maximum mass, $m_\mathrm{max}$.  All values of $m_\mathrm{max}$ produce the same upper bound. \label{fig:lam-mass}}
\end{figure}

\textit{The common equation of state constraint.}---To explore imposing a common EOS constraint, we employ a piecewise polytrope scheme ~\cite{Lattimer:2015nhk} to simulate thousands of equations of state.  Each EOS obeys causality, connects at low densities to the well-known EOS of neutron star crusts~\cite{Lattimer:2012nd}, is constrained by experimental and theoretical studies of the symmetry properties of matter near the nuclear saturation density, and satisfies the observational constraint for the maximum mass of a neutron star, $m_\mathrm{max}\ge2M_\odot$~\cite{Antoniadis:2013pzd}. Figure~\ref{fig:lam-mass} shows the results of Tolman-Oppenheimer-Volkoff (TOV) integrations~\cite{Oppenheimer:1939ne,Postnikov:2010yn} to determine $\Lambda$ as functions of $m$, $R$, and the EOS. Each configuration is color coded according to its
radius. In the relevant mass range, $\Lambda$ generally varies as
$m^{-6}$. For a given mass $m$, there is an inherent spread of about a factor of ten in $\Lambda$, which is correlated with $R^6$. We find that the star's tidal deformability is related to its compactness parameter 
$\beta=Gm/(Rc^2)$ by the relation $\Lambda~\simeq a\beta^{-6}$. We find that $a=0.0093\pm0.0007$ bounds this relation if $1.1M_\odot\le m\le1.6M_\odot$
(note that this is a bound, not a confidence interval). The additional power of $\beta^{-1}$ in the $\Lambda-\beta$ relation, relative to $\beta^{-5}$ in Eq.~(\ref{eq:lambda_12}), originates because the dimensionless tidal Love number, $k_2$, varies roughly as $\beta^{-1}$ for masses $\geq$
$1M_\odot$, although this is not the case for all masses~\cite{Postnikov:2010yn}. For $m\to0$ we see that $k_2\to0$ so that $k_2$ is proportional to $\beta$ with a positive power, but since neutron stars with $m < 1\,M_\odot$ are physically unrealistic, that domain is not pertinent to this Letter.

We observed that, for nearly every specific EOS, the range of stellar radii in the mass range of interest for GW170817 is typically small. As long as $m_{\rm max}\ge 2M_\odot$, the piecewise polytrope study reveals $\langle \Delta R \rangle =-0.070$ km and $\sqrt{\langle(\Delta R)^2\rangle}~=0.11$ km, where $\Delta R \equiv R_{1.6}-R_{1.1}$ with $R_{1.1,1.6}$ the radii of stars with $m=1.1$ and $m=1.6M_\odot$, respectively. Therefore, for masses relevant for GW170817, each EOS assigns a common value of $\hat R$ to stellar radii with little sensitivity to the mass. We can combine the relations $\Lambda~\simeq a\beta^{-6}$ and $R_1=R_2$ to find the simple prescription $\Lambda_1=q^6\Lambda_2$. We impose the common EOS constraint in our analysis using this relation. The exponent of $q$ changes with chirp mass $\mathcal{M}$ and for $\mathcal{M} > 1.5\,M_\odot$ this relation has to be modified. However, this is not relevant for the study of GW170817.

\textit{Implications for the neutron star radius.}---The common EOS constraint allows us to show that the binary tidal deformability $\tilde\Lambda$ is essentially a function of the chirp mass ${\cal{M}}$, the common radius $\hat R$, and the mass ratio $q$, but that its dependence on $q$ is very weak. Substituting the expressions $\Lambda~\simeq a\beta^{-6}$ and $R=\hat R$ into Eq. (\ref{eq:lambda_t0}), we find
\begin{equation}
\tilde{\Lambda}=
\frac{16a}{13}
\left(
\frac{\hat{R}c^2}{G{\cal M}}
\right)^6 f(q).
\label{eq:lambda_t1}\end{equation}
where $f(q)$ is very weakly dependent on $q$:
\begin{equation}
f(q)=q^{8/5}(12-11q+12q^2)(1+q)^{-26/5}.
\end{equation}
For example, if we compare a binary with $q = 0.75$ to an equal mass binary, we find $f(0.75)/f(1)=1.021$. As long as $q\ge0.6$, valid for $1M_\odot\le m\le 1.6 M_\odot$ for both stars, we infer from Eq.~(\ref{eq:lambda_t1}),
\begin{equation}
\tilde{\Lambda}=a^\prime\left(\frac{\hat{R}c^2}{G{\cal M}}\right)^6,
\label{eq:lambda_t2}\end{equation}
where $a^\prime=0.0042\pm0.0004$. The Supplemental Material~\cite{supp} shows TOV integrations for a range of EOS that validate this relationship. For stars with masses comparable to GW170817, the common radius $\hat R$ can be found from the inversion of Eq. (\ref{eq:lambda_t2}),
\begin{equation}
\hat R\simeq R_{1.4}\simeq (11.2\pm0.2)\frac{\cal M}{M_\odot}\left(\frac{\tilde{\Lambda}}{800}\right)^{1/6}{\rm~km}.
\label{eq:rhat}\end{equation}
The quoted errors originate from the uncertainties in $a$ and $q$, and amount, in total, to 2\%.

\textit{Parameter estimation methods.}---
We use Bayesian inference to measure the parameters of GW170817 \cite{Christensen:2001cr}. We calculate the posterior probability density function, $p(\vec{\theta}|\vec{d}(t),H)$, for the set of parameters $\vec{\theta}$ for the gravitational-waveform model, $H$, given the LIGO Hanford, LIGO Livingston, and Virgo data $\vec{d}(t)$~\cite{Vallisneri:2014vxa,gw170817-losc}
\begin{equation}
p(\vec{\theta}|\vec{d}(t),H) = \frac{p(\vec{\theta}|H) p(\vec{d}(t)|\vec{\theta},H)}{p(\vec{d}(t)|H)}.
\label{eq:postpdf}
\end{equation}
The prior, $p(\vec{\theta}|H)$, is the set of assumed probability distributions for the waveform parameters. The likelihood $p(\vec{d}(t)|\vec{\theta},H)$ assumes a Gaussian model for the detector noise~\cite{Rover:2006bb}. Marginalization of the likelihood to obtain the posterior probabilities is performed using Markov Chain Monte Carlo (MCMC) techniques using the PyCBC inference software \cite{Biwer:2018osg,alex_nitz_2018_1208115} and the parallel-tempered EMCEE sampler \cite{emcee,vousden:2016,mcmc}. We fix the sky location and distance to GW170817~\cite{Soares-Santos:2017lru,Cantiello:2018ffy} and calculate the posterior probabilities for the remaining source parameters. Following Ref.~\cite{TheLIGOScientific:2017qsa}, the waveform model $H$ is the restricted TaylorF2 post-Newtonian aligned-spin model \cite{Sathyaprakash:1991mt,Buonanno:2009zt,Arun:2008kb,Mikoczi:2005dn,Bohe:2013cla,Vines:2011ud}.
Technical details of our parameter estimation and a comparison to Fig.~5 of Ref~\cite{TheLIGOScientific:2017qsa} are provided as Supplemental Material~\cite{supp}.

To implement the common EOS constraint we construct the priors on $\Lambda_{1,2}$ according to
\begin{equation}
\Lambda_1=q^3\Lambda_s,\qquad\Lambda_2=q^{-3}\Lambda_s,
\label{eq:lambdas}\end{equation}
where $\Lambda_s \sim U[0,5000]$. We discard draws with 
$\tilde{\Lambda} > 5000$, since these values are beyond the range of all plausible EOS. The resulting prior on $\tilde\Lambda$ is uniform between $0$ and $5000$.  We also perform analyses that do not assume the common EOS constraint where we allow completely uncorrelated priors for $\Lambda_{1,2}$. This allows us to compare the evidences between these hypotheses. For the uncorrelated $\Lambda_{1,2}$ analyses, the prior for $\Lambda_1 \sim U[0,1000]$ and $\Lambda_2 \sim U[0,5000]$ with these intervals set by the range of plausible equations of state in the mass range of interest, our convention of $m_1 \geq m_2$, and discarding draws with $\tilde\Lambda > 5000$.

\begin{figure*}[t]
  \includegraphics[width=\textwidth]{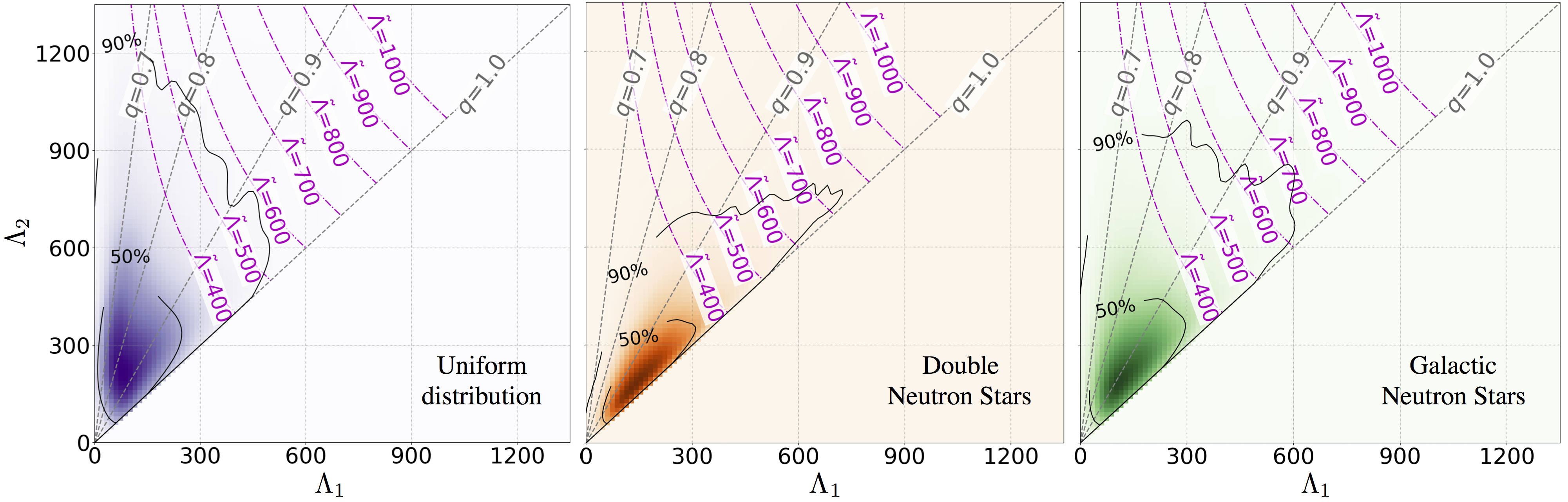}
  \caption{Posterior probability densities for $\Lambda_{1,2}$ with the common EOS constraint using uniform (left), double neutron stars (middle), and Galactic neutron stars (right) component mass priors. The 50$\%$ and 90$\%$ credible region contours are shown as solid curves. Overlaid are contours of $\tilde{\Lambda}$ (in magenta) and $q$ (in gray). The values of $\Lambda_1$ and $\Lambda_2$ forbidden by causality have been excluded from the posteriors.
\label{fig:lamb12_1abc}
\vspace*{-0.3cm}%
}
\end{figure*}
The choice of mass prior can have an impact on the recovery of the tidal deformability~\cite{Agathos:2015uaa}. To investigate this, we carry out our parameter estimation analyses using three different priors on the binary's component masses. First, we assume a uniform prior on each star's mass, with $m_{1,2} \sim U[1,2]\, M_\odot$. Then, we assume a Gaussian prior on the component masses $m_{1,2} \sim N(\mu = 1.33, \sigma = 0.09)\, M_\odot$, which is a fit to masses of neutron stars observed in  double neutron star systems~\cite{Ozel:2016oaf}. The third prior assumes that the component masses are drawn from a fit to the observed mass distributions of recycled and slow pulsars in the Galaxy with $m_1 \sim N(\mu = 1.54, \sigma = 0.23)\, M_\odot$ and $m_2 \sim N(\mu = 1.49, \sigma = 0.19)\, M_\odot$~\cite{Ozel:2016oaf}. We impose the constraint $m_1 \geq m_2$ which leads to $\Lambda_2 \geq \Lambda_1$. For all our analyses, the prior on the component spins is $\chi_{1,2} \sim U[-0.05,0.05]$, consistent with the expected spins of field binaries when they enter the LIGO-Virgo sensitive band~\cite{Brown:2012qf}.

\textit{Results.}---We perform parameter estimation for each mass prior with and without the common EOS constraint and calculate the Bayes factor---the ratio of the evidences $p(\vec{d}(t)|H)$---between the common EOS constrained and unconstrained analyses. We find Bayes factors $\mathcal{B}$ of 369, 125, and 612 for the three mass priors, respectively, indicating that the data strongly favor the common EOS constraint in all cases. The full posterior probability densities of the parameters $p(\vec{\theta}|\vec{d}(t),H)$ for the common EOS runs are shown in the Supplemental Material~\cite{supp} and are available for download at Ref.~\cite{gw170817commoneos}. 
Figure \ref{fig:lamb12_1abc} shows the posterior probability densities for $\Lambda_1$ and $\Lambda_2$ with 90\% and 50\% credible region contours. Overlaid are $q$ contours and $\tilde{\Lambda}$ contours obtained from Eq.~(\ref{eq:lambda_t0}), $\Lambda~\simeq a\beta^{-6}$, and $R_1 \simeq R_2 \simeq \large \hat R$ as
\begin{equation}
\Lambda_1(\tilde{\Lambda},q)={\frac{13}{16}}\tilde{\Lambda}{\frac{q^2(1+q)^4}{12q^2-11q+12}},
\label{eq:l1l2}\quad \Lambda_2(\tilde{\Lambda},q) = q^{-6} \Lambda_1 .\end{equation}
Because of our constraint $\Lambda_2 \geq \Lambda_1$, our credible contours are confined to the region where $q \leq 1$. One can easily demonstrate that $\Lambda_2 \geq \Lambda_1$ is valid unless $(c^2/G)dR/dm > 1$, which is impossible for realistic equations of state. For the entire set of piecewise polytropes satisfying $m_\mathrm{max}>2M_\odot$ we considered, $(c^2/G)dR/dm$ never exceeded 0.26. Even if a first order phase transition appeared in stars with masses between $m_2$ and $m_1$, it would necessarily be true that $dR/dm < 0$ across the transition. Because of the $q$ dependence of $\Lambda_1$, $\Lambda_2$, the credible region enclosed by the contours broadens from the double neutron star (most restricted), to the pulsar, to the uniform mass (least restricted) priors. However, the upper bound of the credible region is robust.

We find $\tilde{\Lambda}=
205^{+415}_{-167}$ for the uniform component mass prior, $\tilde{\Lambda}=234^{+452}_{-180}$ for the prior informed by double neutron star binaries in the Galaxy, and $\tilde{\Lambda}=218^{+445}_{-173}$ for the prior informed by all Galactic neutron star masses (errors represent 90\% credible intervals). Our measurement of $\tilde{\Lambda}$ appears to be robust to the choice of component mass prior, within the (relatively large) statistical errors on its measurement. The Bayes factors comparing the evidence from the three mass priors are of order unity, so we cannot claim any preference between the mass priors.

\begin{figure}[t]
  \includegraphics[width=\columnwidth]{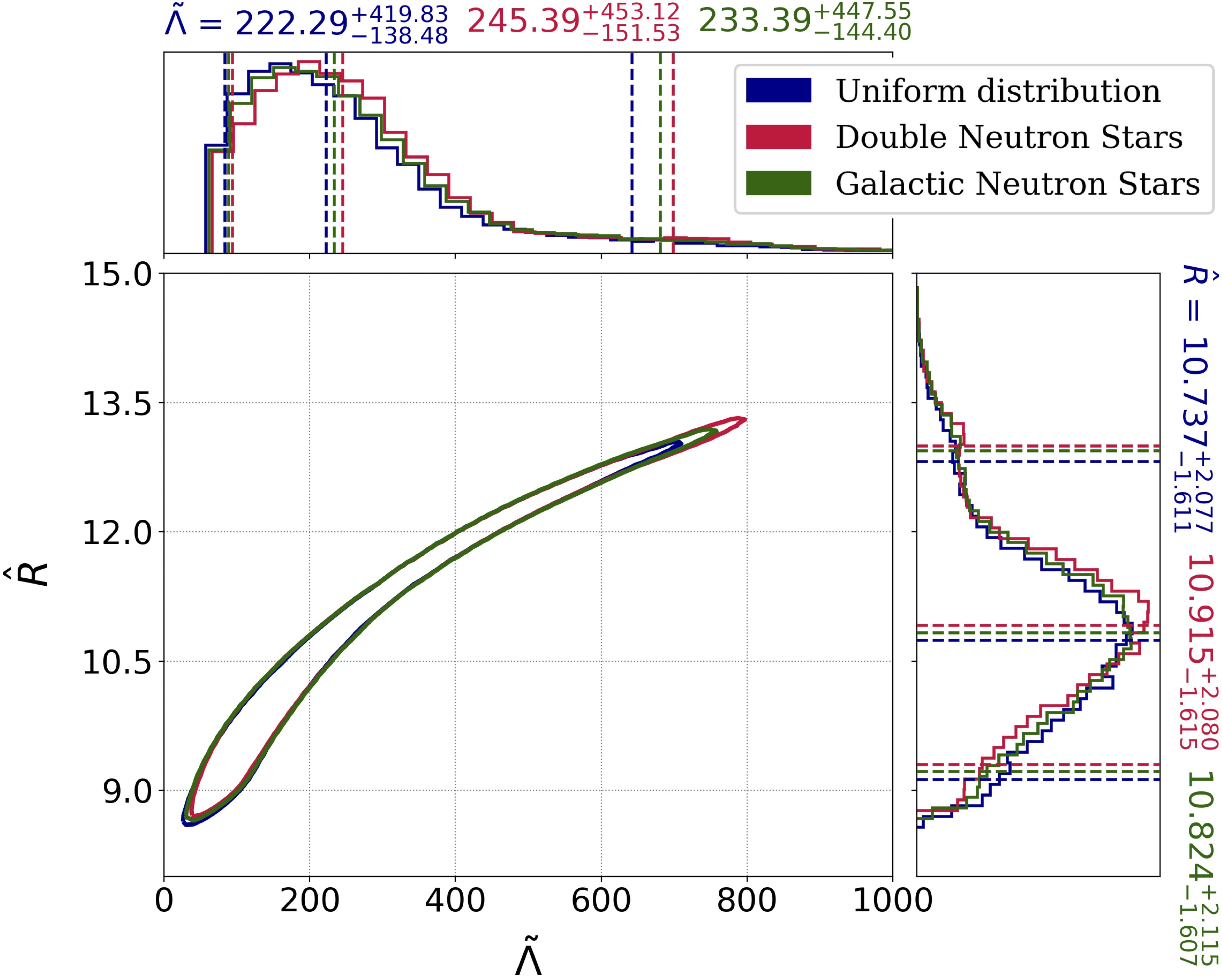}
  \caption{The 90\% credible region of the posterior probability for the common radius $\hat{R}$ and binary tidal deformability $\tilde\Lambda$ with the common EOS constraint for the three mass priors. The posteriors for the individual parameters are shown with dotted lines at  the 5$\%$, 50$\%$ and 95$\%$ percentiles. The values of $\tilde\Lambda$, and hence $\hat{R}$ forbidden by causality have been excluded from the posteriors. 
\label{fig:radius_lambda}%
\vspace*{-0.5cm}%
}
\end{figure}

The 90\% credible intervals on $\tilde{\Lambda}$ obtained from the gravitational-wave observations include regions forbidden by causality. Applying a constraint to our posteriors for the causal lower limit of $\Lambda$ as a function of $m$~\cite{Zhao:2018nyf}, we obtain $\tilde{\Lambda}=
222^{+420}_{-138}$ for the uniform component mass prior, $\tilde{\Lambda}=245^{+453}_{-151}$ for the prior informed by double neutron star binaries in the Galaxy, and $\tilde{\Lambda}=233^{+448}_{-144}$ for the prior informed by all Galactic neutron star masses (errors represent 90\% credible intervals). Using Eq.~(\ref{eq:rhat}), we map our $\cal{M}$ posteriors and $\tilde{\Lambda}$ posteriors (with the causal lower limit applied) to $\hat{R} \simeq R_{1.4}$ posteriors, allowing us to estimate the common radius of the neutron stars for GW170817 for each mass prior. Figure~\ref{fig:radius_lambda} shows the posterior probability distribution for the binary tidal deformation $\tilde\Lambda$ and the common radius $\hat{R}$ of the neutron stars in the binary. Our results suggest a radius $\hat R = 10.7^{+2.1}_{-1.6} \pm 0.2$ km (90\% credible interval, statistical and systematic errors) for the uniform mass prior,  $\hat R = 10.9^{+2.1}_{-1.6} \pm 0.2$ km for double neutron star mass prior, and $\hat R = 10.8^{+2.1}_{-1.6} \pm 0.2$ km for the prior based on all neutron star masses.

For the uniform mass prior, we computed the Bayes factor comparing a model with a prior $\Lambda_s \sim U[0,5000]$ to a model with a prior $\Lambda_s \sim U[0,100]$. We find $\log_{10}(\mathcal{B}) \sim 1$, suggesting that the data favors a model that includes measurement of tidal deformability $\tilde\Lambda \gtrsim 100$. However, the evidences were calculated using thermodynamic integration of the MCMC chains~\cite{emcee}. We will investigate model selection using, e.g., nested sampling \cite{skilling2006} in a future work.

Finally, we note the post-Newtonian waveform family used will result in systematic errors in our measurement of the tidal deformability \cite{Wade:2014vqa,Lackey:2014fwa}. However, this waveform family allows a direct comparison to the results of Ref.~\cite{TheLIGOScientific:2017qsa}. Accurate modeling of the waveform is challenging, as the errors in numerical simulations are comparable to the size of the matter effects that we are trying to measure~\cite{Barkett:2015wia}. Waveform systematics and comparison of other waveform models (e.g., \cite{Bernuzzi:2014owa}) will be investigated in a future work.

\textit{Discussion.}---Using Bayesian parameter estimation, we have measured the tidal deformability and common radius of the neutron stars in GW170817. Table~\ref{tab:summary_table} summarizes our findings.  To compare to Ref.~\cite{TheLIGOScientific:2017qsa}, which reports a 90\% upper limit on $\tilde\Lambda \le 800$ under the assumption of a uniform prior on $\tilde\Lambda$, we integrate the posterior for $\tilde{\Lambda}$ to obtain 90\% upper limits on $\tilde{\Lambda}$.   For the common EOS analyses, these are $485$, $521$, and $516$ for the uniform, double neutron star, and Galactic neutron star component mass priors, respectively. We find that, in comparison to the unconstrained analysis, the common EOS assumption significantly reduces the median value and 90\% confidence upper bound of $\tilde\Lambda$ by about 28\% and 19\%, respectively, for all three mass priors. The difference between our common EOS results for the three mass priors is consistent with the physics of the gravitational waveform. At constant $\mathcal{M}$, decreasing $q$ causes the binary to inspiral more quickly \cite{Hannam:2013uu}. At constant $\mathcal{M}$ and constant $q$, increasing $\tilde\Lambda$ also causes the binary to inspiral more quickly, so there is a mild degeneracy between $q$ and $\tilde\Lambda$. The uniform mass prior allows the largest range of mass ratios, so we can fit the data with a larger $q$ and smaller $\tilde\Lambda$. The double neutron star mass prior allows the smallest range of mass ratios, and so, a larger $\tilde\Lambda$ is required to fit the data, with the Galactic neutron star mass prior lying between these two cases. 

\begin{table}[t]\label{tab:parameters}
\setlength{\tabcolsep}{3.8pt}
\centering\begin{tabular}{lcccc} 
\hline
\rule{0pt}{3ex}%
Mass prior \quad & \quad $\tilde{\Lambda}$ \quad & \quad $\hat{R}$ (km) \quad & \quad $\mathcal{B}$ \quad & \quad $\tilde{\Lambda}_{90\%}$\quad \\\hline
\rule{0pt}{3ex}%
Uniform & 222$^{+420}_{-138}$ & 10.7$^{+2.1}_{-1.6}\pm 0.2$ & 369 & $< 485$\\
Double neutron star & 245$^{+453}_{-151}$ & 10.9$^{+2.1}_{-1.6}\pm 0.2$ &  125 & $< 521$ \\
Galactic neutron star & 
233$^{+448}_{-144}$ & 10.8$^{+2.1}_{-1.6}\pm 0.2$ & 612 & $< 516$ \\
\hline
\end{tabular}
\caption{Results from parameter estimation analyses using three different mass prior choices with the common EOS constraint, and applying the causal minimum constraint to $\Lambda(m)$. We show 90$\%$ credible intervals for $\tilde{\Lambda}$, 90$\%$ credible intervals and systematic errors for $\hat{R}$, Bayes factors $\mathcal{B}$ comparing our common EOS to the unconstrained results, and the 90\% upper limits on $\tilde\Lambda$.%
\vspace*{-0.5cm}%
}
\label{tab:summary_table}
\end{table}
Nevertheless, considering all analyses we performed with different mass prior choices, we find a relatively robust measurement of the common neutron star radius with a mean value $\langle \large \hat R \rangle$ = 10.8 km bounded above by $\hat R < 13.2$~km and below by $\hat R > 8.9$~km. Nuclear theory and experiment currently predict a somewhat smaller range by 2 km but with approximately the same centroid as our results~\cite{Lattimer:2012nd,Lattimer:2012xj}. A minimum radius 10.5--11 km is strongly supported by neutron matter theory~\cite{Gandolfi:2011xu,Lynn:2015jua,Drischler:2015eba}, the unitary gas~\cite{Kolomeitsev:2016sjl}, and most nuclear experiments~\cite{Lattimer:2012nd,Lattimer:2012xj,Tews:2012fj}. The only major nuclear experiment that could indicate radii much larger than 13 km is the PREX neutron skin measurement, but this has published error bars much larger than previous analyses based on antiproton data, charge radii of mirror nuclei, and dipole resonances. 
Our results are consistent with photospheric radius expansion measurements of x-ray binaries which obtain $R \approx 10$--$12$~km~\cite{Ozel:2016oaf,Steiner:2010fz,Degenaar:2018lle}. Reference~\cite{Guillot:2014lla} found from an analysis of five neutron stars in quiescent low-mass x-ray binaries a common neutron star radius $9.4\pm1.2$ km, but systematic effects including uncertainties in interstellar absorption and the neutron stars' atmospheric compositions are large. Other analyses have inferred $12\pm0.7$~\cite{Lattimer:2013hma} and $12.3\pm1.8$ km~\cite{Shaw:2018wxh} for the radii of $1.4M_\odot$ quiescent sources. 

We have found that the relation $q^{7.48} < \Lambda_1/\Lambda_2 < q^{5.76}$, in fact, completely bounds the uncertainty for the range of $\mathcal{M}$ relevant to GW170817, assuming $m_2 > 1M_\odot$ \cite{Zhao:2018nyf} and that no strong first-order phase transitions occur near the nuclear saturation density (i.e., the case in which $m_1$ is a hybrid star and $m_2$ is not). Analyses using this prescription instead of the $q^6$ correlation produce insignificant differences in our results \cite{De:2018prep}. Since models with the common EOS assumption are highly favored over those without this assumption, our results support the absence of a strong first-order phase transition in this mass range.

In this Letter, we have shown that, for binary neutron star mergers consistent with observed double neutron star systems~\cite{Tauris:2017omb}, assuming a common EOS implies that $\Lambda_1 / \Lambda_2 \simeq q^6$. We find evidence from GW170817 that favors the common EOS interpretation compared to uncorrelated deformabilities. Although previous studies have suggested that measurement of the tidal deformability is sensitive to the choice of mass prior \cite{Agathos:2015uaa}, we find that varying the mass priors does not significantly influence our conclusions suggesting that our results are robust to the choice of mass prior. Our results support the conclusion that we find the first evidence for finite size effects using gravitational-wave observations. 

Recently, the LIGO/Virgo collaborations have placed new constraints on the radii of the neutron stars using GW170817~\cite{Abbott:2018exr}. The most direct comparison is between our uniform mass prior result ($\hat R = 10.7^{+2.1}_{-1.6} \pm 0.2$) and the LIGO/Virgo method that uses equation-of-state-insensitive relations~\cite{Yagi:2016qmr,Chatziioannou:2018vzf} ($R_1 = 10.8^{+2.0}_{-1.7}$ and $R_2 = 10.7^{+2.1}_{-1.5}$~km). This result validates our approximation $R_1=R_2$ used to motivate the prescription $\Lambda_1=q^6\Lambda_2$, and Eqs.~(\ref{eq:lambda_t1}, \ref{eq:lambda_t2}). Our statistical errors are comparable to the error reported by LIGO/Virgo. Systematic errors from EOS physics of $\pm 0.2$~km are added as conservative bounds to our statistical errors, broadening our measurement error, whereas Ref.~\cite{Abbott:2018exr} marginalized over these errors in the analysis. Reference~\cite{Abbott:2018exr} also investigates a method of directly measuring the parameters of the EOS which results in smaller measurement errors. Investigation of these differences between our analysis and the latter approach will be pursued in a future paper. 

Observations of future binary neutron star mergers will allow further constraints to be placed on the deformability and radius, especially if these binaries have chirp masses similar to GW170817 as radio observations suggest. As more observations improve our knowledge of the neutron star mass distribution, more precise mass-deformability correlations can be used to further constrain the star's radius.

We thank Stefan Ballmer, Swetha Bhagwat, Steven Reyes, Andrew Steiner, and Douglas Swesty for helpful discussions. We particularly thank Collin Capano and Alexander Nitz for contributing to the development of PyCBC Inference. This Letter was supported by NSF Grants No. PHY-1404395 (D.A.B., C.M.B.), No. PHY-1707954 (D.A.B., S.D.), No. PHY-1607169 (S.D.), No. AST-1559694 (D.F.), No. AST-1714498 (E.B.), and DOE Award No. DE-FG02-87ER40317 (J.M.L.). Computations were supported by Syracuse University and NSF Grant No. OAC-1541396. D.A.B., E.B., S.D., and J.M.L. thank the Kavli Institute for Theoretical Physics which is supported by the NSF Grant No. PHY-1748958. The gravitational-wave data used in this Letter was obtained from the LIGO Open Science Center.


%

\end{bibunit}
\clearpage
\newpage
\begin{bibunit}[supp]
\setcounter{figure}{0}
\setcounter{equation}{0}
\setcounter{page}{1}
\onecolumngrid
\begin{center}
{\large \bf Supplemental Material}
\vspace*{1cm}
\twocolumngrid
\end{center}

\input{supp}

\end{bibunit}
\clearpage
\newpage
\begin{bibunit}[erratum]
\setcounter{figure}{0}
\setcounter{equation}{0}
\setcounter{page}{1}
\onecolumngrid
\begin{center}
{\large \bf Erratum}
\vspace*{1cm}
\twocolumngrid
\end{center}

\input{erratum}
\end{bibunit}

\end{document}

%% file: supp.tex
\textit{The common neutron star radius---}To validate the relationship
\begin{equation}
\tilde{\Lambda}=a^\prime\left(\frac{\hat{R}c^2}{G{\cal M}}\right)^6,
\label{eq:lambda_t2_supp}\end{equation}
where $a^\prime=0.0042\pm0.0004$, we perform Tolman-Oppenheimer-Volkoff (TOV) integrations~\cite{Oppenheimer:1939ne} as described in the main text.
\begin{figure}[b]
  \includegraphics[width=\columnwidth]{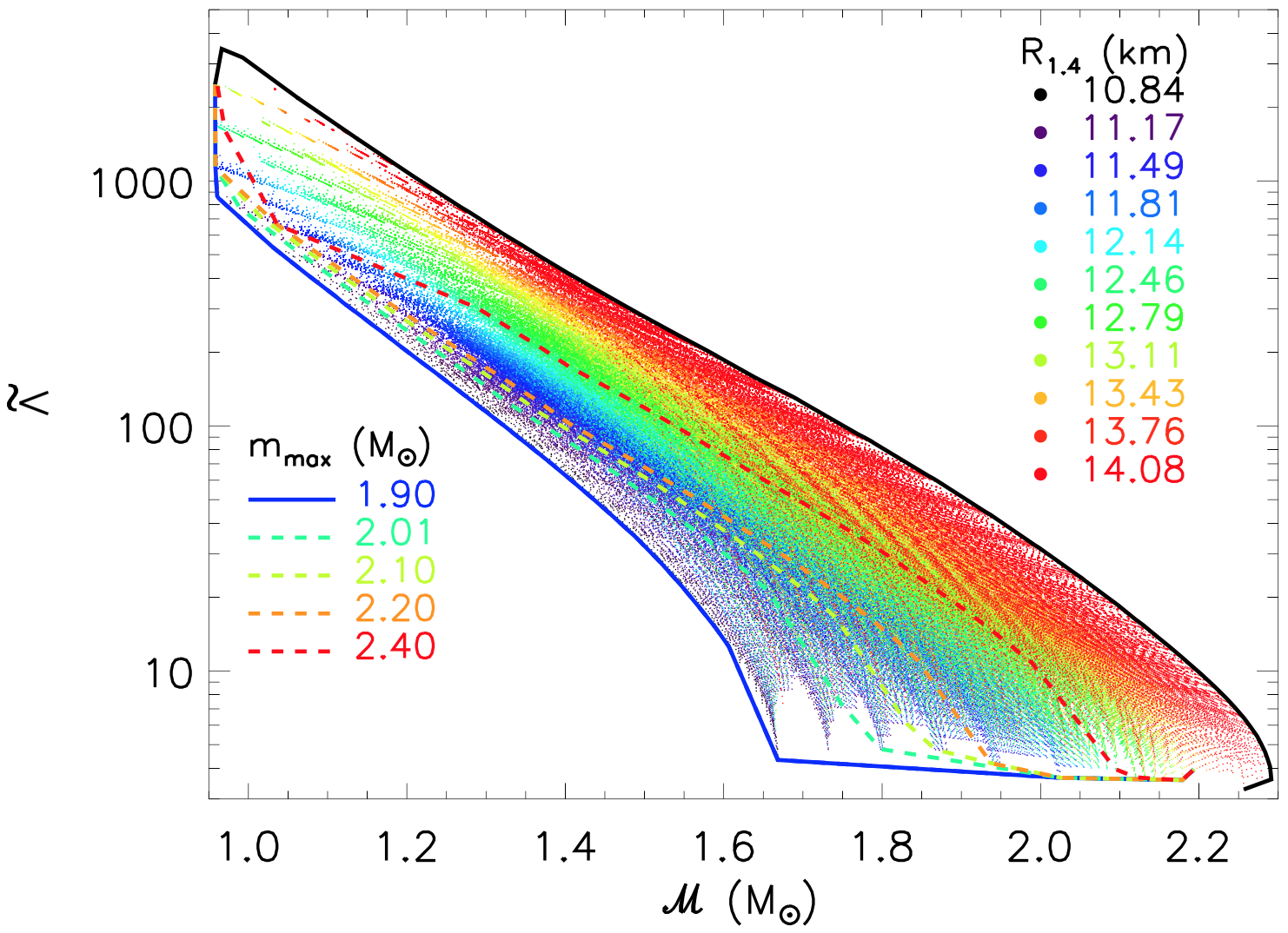}
  \caption{The dimensionless binary tidal deformability $\tilde \Lambda$ as a function of chirp mass $\cal{M}$ for piecewise polytropes with parameters bounded by causality, neutron matter studies and nuclear experiments. Various binary mass combinations with each equation of state result in many points which are color-coded according to that equation of state's value of $R_{1.4}$.  The values of $\tilde\Lambda(\cal{M})$ are bounded according to the assumed value of the maximum neutron star mass, $m_\mathrm{max}$. \label{fig:lambar-mass}}
\end{figure}
The relationship between $\tilde\Lambda$, chirp mass ${\cal M}$ and the common radius $\hat R$ closely resembles the relation between $\Lambda$, the neutron star mass $m$ and radius $R$. We confirm this using piecewise polytropes as shown in Fig.~\ref{fig:lambar-mass}, which was prepared using the results computed to create Fig.~1 in the main paper. A single mass-radius curve is generated for each equation of state containing $N$ masses between $1M_\odot$ and $m_\mathrm{max}$ for that EOS. $N(N-1)$ values of $\tilde\Lambda$ and $\cal{M}$ are then computed for all the unique combinations of $m_1$ and $m_2$ from these $N$ masses. The resulting points are plotted in Fig.~\ref{fig:lambar-mass}, and are color-coded by that equation of state's value of $R_{1.4}$. The process is repeated for all combinations of the parameters controlling the piecewise polytropic EOS described in \cite{Lattimer:2015nhk}. For the entries bounded by $0.9 M_\odot\le~{\cal M} \le~1.3M_\odot$, an interval including GW170817, Eq. (\ref{eq:lambda_t2_supp}) is determined by finding the upper and lower bounds of $a^\prime=\tilde\Lambda[G{\cal M}/(R_{1.4}c^2)]^6$.
\begin{figure}[t]
  \includegraphics[width=\columnwidth]{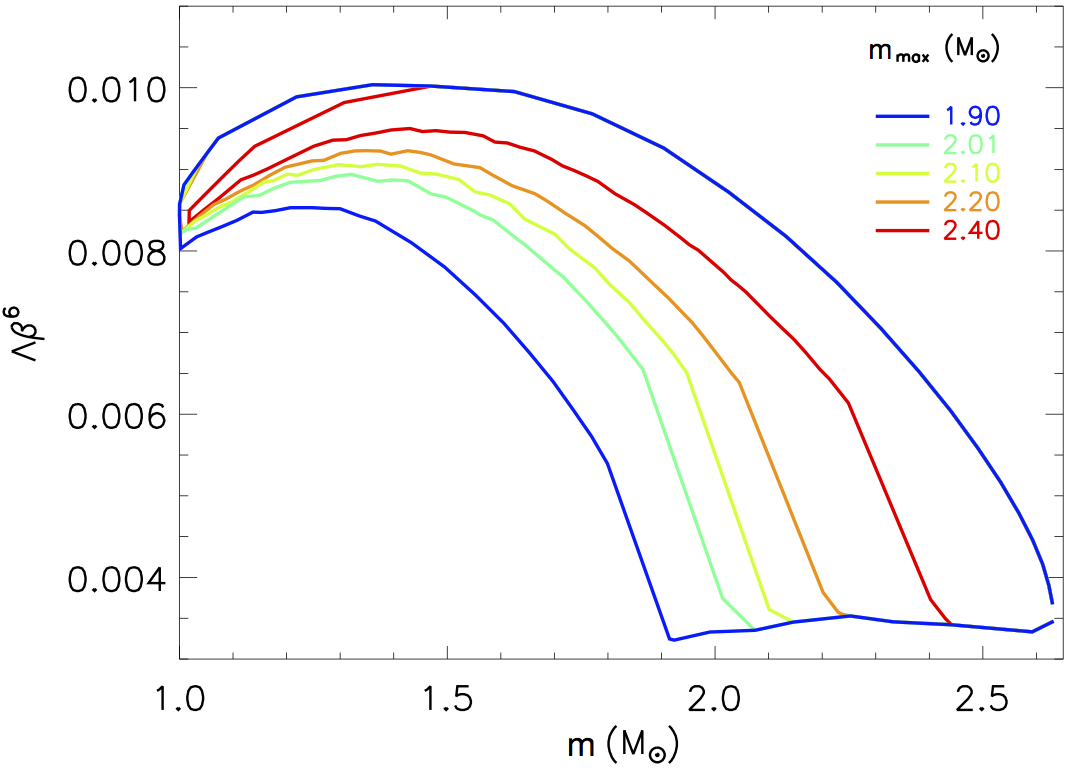}
  \caption{$\Lambda \beta^{6}$ as a function of neutron star mass $m$ for physically realistic polytropes. The curves show lower bounds to $\Lambda$ for a given mass $m$ for different assumed lower limits to the neutron star maximum mass, $m_\mathrm{max}$. The curves are colored by $m_\mathrm{max}$. All values of $m_\mathrm{max}$ produce the same upper bound. \label{fig:lam-beta6}}
\end{figure}

The numerical results also confirmed our value for $a^\prime$. This is valid for values of $\mathcal{M} < 1.3 M_\odot$, which is relevant not only for GW170817 but also for all known double neutron star binaries, which are clustered in the narrow range $1.09M_\odot<{\cal M}<1.25M_\odot$ \cite{Tauris:2017omb,Ozel:2016oaf,Lattimer:2012nd}. The robustness of $\tilde{\Lambda}\propto\beta^{-6}$ confirms that the assumption $R_1=R_2$ is a valid proposition. To justify the degree of correlation in the $\Lambda~\simeq a\beta^{-6}$ that we established in the main text, we show in Fig.~\ref{fig:lam-beta6} the dependence of $\Lambda \beta^{6}$ on $m$. The variation of $\Lambda \beta^{6}$ is negligible in the mass range relevant for GW170817, $1.1 < m < 1.6 M_\odot$, thus confirming the validity of the $\Lambda~\simeq a\beta^{-6}$ relation for GW170817.

\textit{Causal constraints on the tidal deformation---}The posterior probability distribution that we observe for the star's tidal deformability includes regions forbidden by causality~\cite{Zhao:2018nyf}. In our results for $\Lambda$ and (and hence $\hat{R}$), we apply the causal lower limit constraint on the tidal deformability $\Lambda$ as a function of mass $m$. We implement this constraint using the following relation, which is valid for $0.4 < m/m_{max} < 0.95$,
\begin{equation}
\begin{split}
\ln\Lambda_{\mathrm min} = 13.42 - 23.01\left(\frac{m}{m_{\mathrm max}}\right) \\
+ 20.53 \left(\frac{m}{m_{\mathrm max}}\right)^2 \\
- 9.599\left(\frac{m}{m_{\mathrm max}}\right)^3.
\end{split}
\end{equation}
Here, we use $m_{\mathrm max} = 2 M_\odot$. Note that $m_{\mathrm max} > 2$, would increase the lower limit for $\Lambda(m)$, so $m_{\mathrm max}=2$ is a conservative choice~\cite{Zhao:2018nyf}.

\textit{Parameter estimation methods}---To measure the source parameters for GW170817, we performed parameter estimation on the Advanced LIGO-Virgo data available at the LIGO Open Science Center \cite{Vallisneri:2014vxa,gw170817-losc}. 
Our analysis was performed with the \textit{PyCBC Inference} software \cite{Biwer:2018osg,alex_nitz_2018_1208115} and the parallel-tempered \textit{emcee} sampler \cite{emcee,vousden:2016} for sampling over the parameter space using Markov Chain Monte Carlo (MCMC) techniques \cite{mcmc}. 

The LOSC data files include a post-processing noise subtraction performed by the LIGO-Virgo Collaboration \cite{gw170817-losc,gw170817-noise}. The LOSC documentation states that these data have been truncated to remove tapering effects due to the cleaning process \cite{gw170817-losc}, however the LOSC data shows evidence of tapering after GPS time $1187008900$ in the LIGO Hanford detector. To avoid any contamination of our results we do not use any data after GPS time $1187008891$. The power spectral density (PSD) used to construct the likelihood was calculated using Welch's method \cite{1161901} with 16~second Hann-windowed segments (overlapped by 8~s) taken from GPS time $1187007048$ to $1187008680$. The PSD estimate is truncated to 8~s length in the time domain using the method described in Ref. \cite{Allen:2005fk}. The gravitational-wave data used in the likelihood is taken from the interval $1187008691$ to $1187008891$. 

\begin{figure}[t]
  \includegraphics[width=\columnwidth]{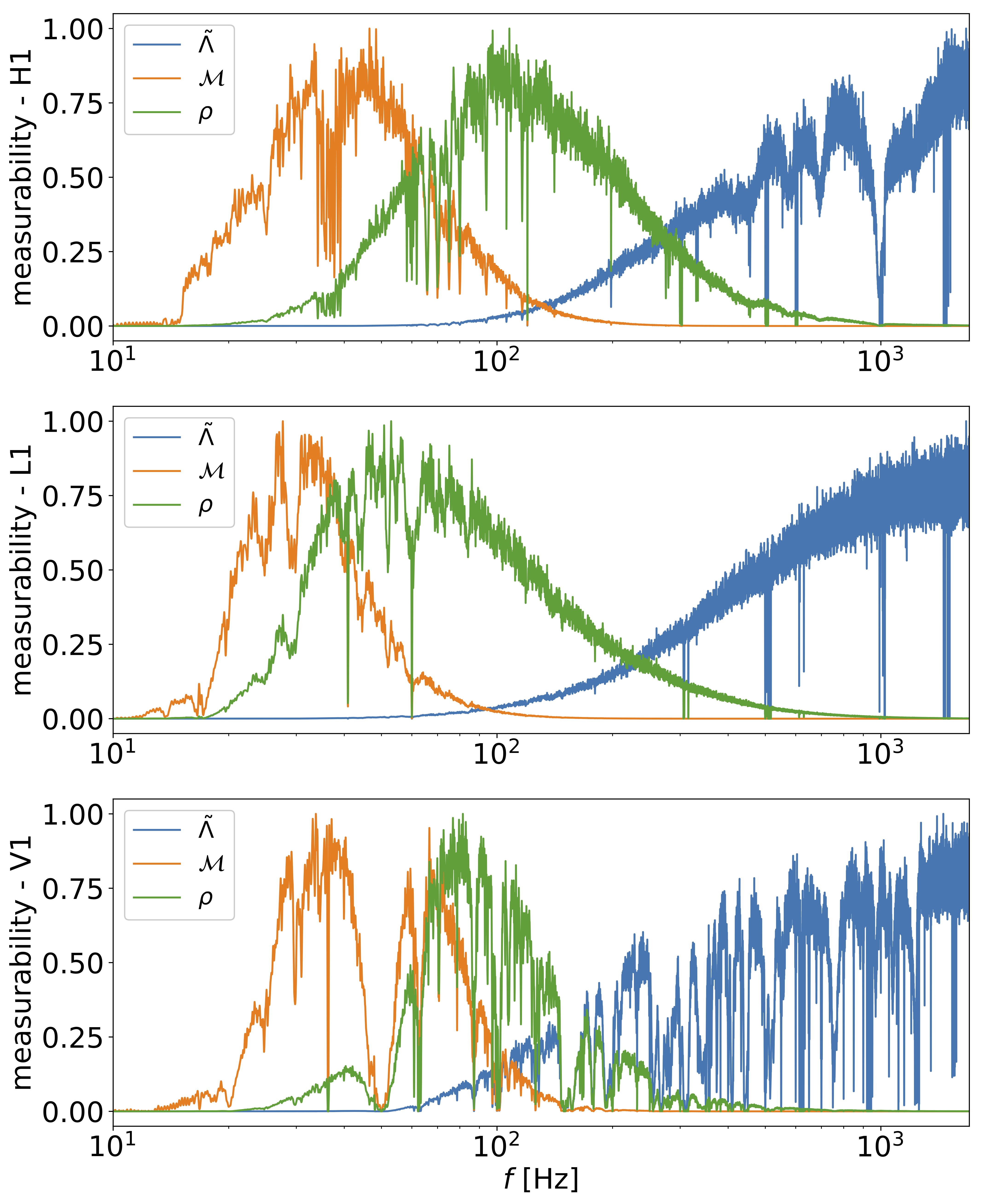}
  \caption{Measurability~\cite{Damour:2012yf} of the chirp mass $\mathcal{M}$, SNR $\rho$ and binary deformability $\tilde{\Lambda}$ in the frequency range 10 Hz - 2000 Hz. Each detector's parameter measurability is scaled to the maximum frequency to show the relative accumulation of measurement over the detector's frequency band. Note that between detectors, L1 is more sensitive than H1, which is more sensitive than V1. Measurability of chirp mass is accumulated primarily at low frequencies, whereas measurability of tidal deformability is accumulated at higher frequencies. We extend computation of the likelihood down to $20$~Hz where the measured signal-to-noise ratio (the logarthim of the likelihood) drops to zero in all three detectors. 
  \label{fig:measurability}}
\end{figure}

Ref.~\cite{Abbott:2018wiz} found that choice of the low-frequency cutoff can have an effect on the measurement of the neutron star tidal deformability and used a different power spectral density estimation technique to that used in our analysis~\cite{Littenberg:2014oda}. We investigated the effect of changing our estimate of the power spectral density with the power spectral density released as supplemental materials to Ref.~\cite{Abbott:2018wiz}. We find that the change in parameter measurements is smaller than the statistical errors, and conclude that the choice of power spectral density estimation technique does not affect our results. To investigate the choice of low-frequency cutoff, we computed the measurabilities of the chirp mass $\mathcal{M}$, signal-to-noise ratio $\rho$, and binary deformability $\tilde{\Lambda}$ in the frequency range 10-2000~Hz. These are defined as the integrand as a function of frequency of the noise moment integrals $I_{\mathrm 10}$, $I_{\mathrm 0}$, and $I_{\mathrm -10}$ (see Ref.~\cite{Damour:2012yf}) and shown in Fig.~\ref{fig:measurability}. It can be seen that the signal-to-noise ratio is non-zero down to a frequency of $\sim$ 20~Hz for all the three detectors. While detector sensitivity at this frequency does not affect the measurability of $\tilde\Lambda$, it does affect the measurability of the chirp mass $\mathcal{M}$. We repeated our analyses at 25~Hz, 23~Hz, and 20~Hz, and found an improvement in the $\mathcal{M}$ measurement when extending until the low-frequency cutoff was 20~Hz. Consequently, we evaluated the likelihood from a low-frequency cutoff of 20~Hz to the Nyquist frequency of 2048~Hz. The improved measurement of $\mathcal{M}$ eliminates regions of higher $\tilde{\Lambda}$ values from the posterior probability densities, and hence better constrains the measurement of this parameter, as shown in Fig~\ref{fig:posterior_overlap}.

The templates for the waveforms used in our parameter estimation analysis are generated using the restricted TaylorF2 waveform model, a Fourier domain waveform model generated using stationary phase approximation. We use the implementation from the LIGO Algorithm Library (LAL)~\cite{lal} accurate to 3.5 post-Newtonian (pN) order in orbital phase \citep{Buonanno:2009zt}, 2.0 pN order in spin-spin, quadrupole-monopole and self-spin interactions\citep{Arun:2008kb,Mikoczi:2005dn}, and 3.5 pN order in spin-orbit interactions \citep{Bohe:2013cla}. The tidal corrections enter at the 5 pN and 6 pN orders~\citep{Vines:2011ud}. The waveforms are terminated at twice the orbital frequency of a test particle at the innermost stable circular orbit of a Schwarzschild black hole of mass $M = m_1 + m_2$, where $m_{1,2}$ are the masses of the binary's component stars. The TaylorF2 model assumes that the spins of the neutron stars are aligned with the orbital angular momentum. Binary neutron stars formed in the field are expected to have small spins, and precession of the binary's orbital plane is not significant~\cite{Brown:2012qf}. 

We fix the sky location of the binary to the right ascension RA =  $197.450374^\circ$ and declination Dec = $-23.381495^\circ$ \cite{Soares-Santos:2017lru} for all of our runs. We also fix the luminosity distance of NGC\,4993 $d_L = 40.7$~Mpc~\cite{Cantiello:2018ffy}. The small error in the known distance of NGC\,4993 produces errors that are much smaller than the errors in measuring the tidal deformability. We have checked that including the uncertainty in the distance error does not affect our conclusions of the tidal deformabilities or radius. The MCMC computes the marginalized posterior probabilities for the remaining source parameters: chirp mass $\mathcal{M}$, mass ratio $q$, the component (aligned) spins $\chi_{1,2} = c J_{1,2}/G m_{1,2}^2$, component tidal deformabilities $\Lambda_{1,2}$, polarization angle $\psi$, inclination angle $\iota$, coalescence phase $\phi_c$, and coalescence time $t_c$. When generating the waveform in the MCMC, each $m_{1, 2}$ draw follows the constraint $m_1 \geq m_2$, and the masses are transformed to the detector frame chirp mass $\mathcal{M}^\mathrm{det}$ and $q$ with a restriction $1.1876 \le \mathcal{M}^\mathrm{det} \le 1.2076$.

\begin{figure}[b]
\includegraphics[width=\columnwidth]{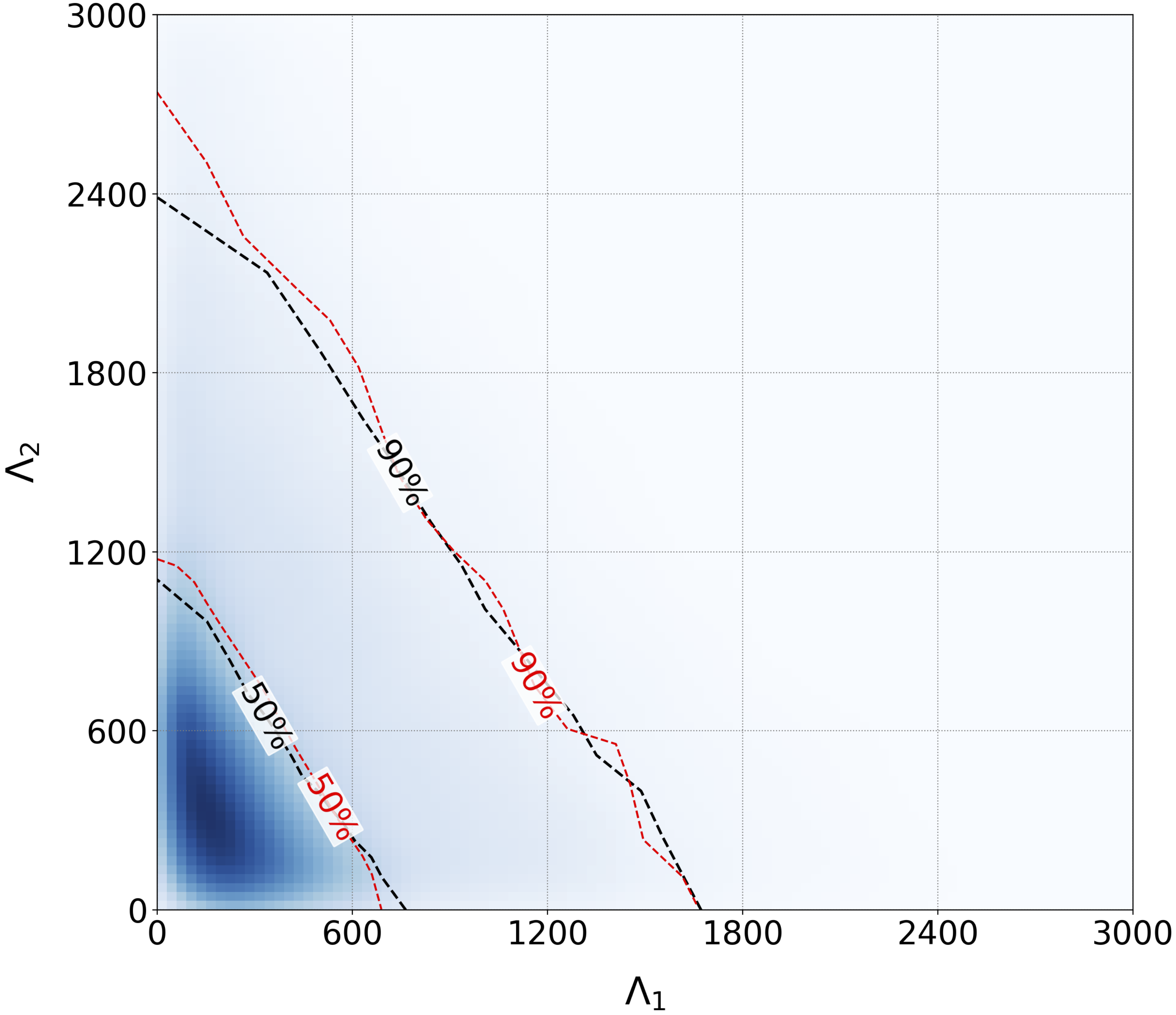}
\caption{Posterior probability density function for $\Lambda_1$, $\Lambda_2$ from unconstrained $\Lambda_{1,2} \sim U[0, 3000]$, $m_{1,2} \sim U[1, 2]$ M$_\odot$, $1.1876 \leq \mathcal{M} \leq 1.2076$, $m_1 \geq m_2$, 30~Hz low-frequency cutoff analysis. The black dotted lines show 50\% and 90\% upper limits from our analysis. The red dotted lines show 50\% and 90\% upper limits from the LIGO-Virgo analysis \cite{TheLIGOScientific:2017qsa}. } 
\label{fig:lv_compare}
\end{figure}

For direct comparison with the results of Ref.~\cite{TheLIGOScientific:2017qsa}, Fig~\ref{fig:lv_compare} shows the posterior probability densities for $\Lambda_{1,2}$ for an MCMC using a $30$~Hz low-frequency cutoff for the uniform component mass prior $m_{1,2} \sim U[1,2]\, M_\odot$, and assuming that the priors on $\Lambda_{1,2}$ are completely uncorrelated ($\Lambda_{1,2} \sim U[0,3000]$). No cut is placed on $\tilde\Lambda$ in this analysis. We have digitized the 50\% and 90\% contours from Fig.~5 of Ref.~\cite{TheLIGOScientific:2017qsa} and compared them to 50\% and 90\% upper limit contours for our result computed using a radial binning to enclose 50\% and 90\% of the posterior probability starting from $\Lambda_1 = \Lambda_2 = 0$. The 90\% contours agree well, with a slight difference in the 50\% contours. Given the accuracy of measuring the tidal deformability, this difference can 
be attributed to small differences in the technical aspects of our analysis compared to that of Ref.~\cite{TheLIGOScientific:2017qsa}. We note that the 90\% confidence contour of Fig.~5 in Ref.~\cite{TheLIGOScientific:2017qsa} with $\Lambda_1=\Lambda_2$, passes through $\tilde\Lambda \approx 1100$. If we impose $\Lambda_1=q^6 \Lambda_2$, then this contour continues to follow $\tilde\Lambda \approx 1100$ for $q \leq 1$. We interpret the difference between this result and the result of Table~I of Ref.~\cite{TheLIGOScientific:2017qsa} $\tilde\Lambda \le 800$ (90\% confidence) as being due to a different choice of prior on $\tilde\Lambda$ (one non-uniform and one uniform).

\begin{figure}[t]
\includegraphics[width=\columnwidth,height=7cm,width=8.5cm]{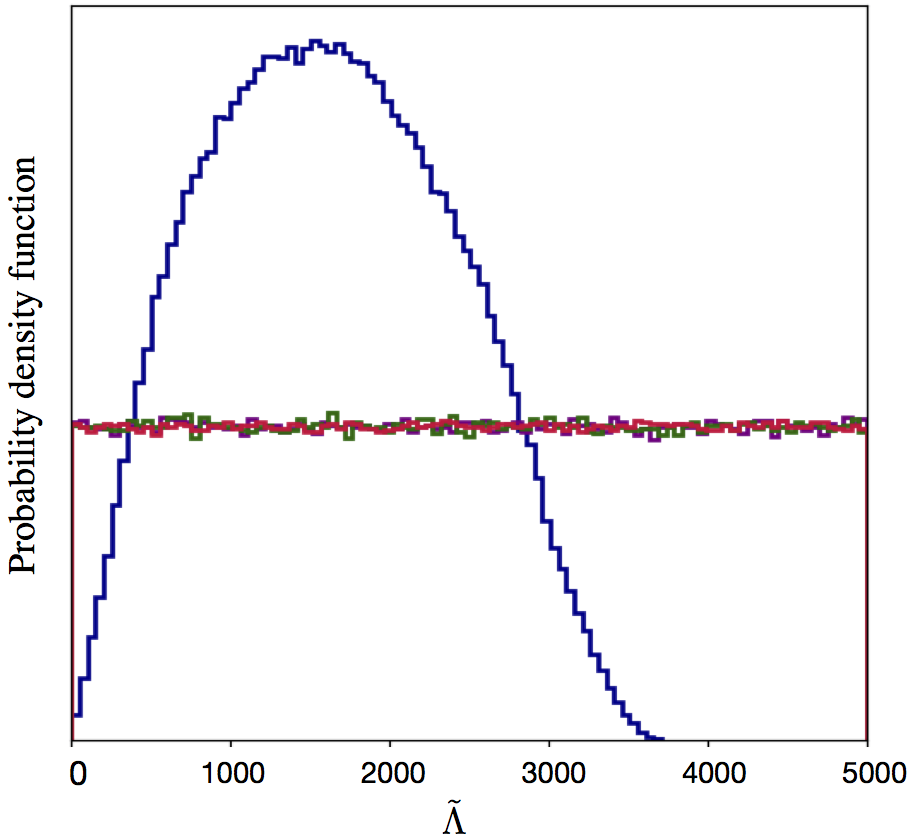}
\caption{Comparison of the prior probability distributions on $\tilde\Lambda$ for the three mass priors imposing the common EOS constraint: uniform (purple), double neutron stars (red), galactic neutron stars (green) with a prior in $\Lambda_{1,2} \sim U[0, 3000]$ and $m_{1,2} \sim U[1, 2] M_\odot$, $m_1 \geq m_2$ without the common EOS constraint (blue). The priors in the common EOS analysis are uniform across the region of interest.} 
\label{fig:lambda_priors}
\end{figure}

Our common equation of state constraint is implemented in the MCMC by drawing a variable $\Lambda_s \sim U[0,5000]$, drawing the component masses from their respective priors and computing
\begin{equation}
\Lambda_1=q^3\Lambda_s,\qquad\Lambda_2=q^{-3}\Lambda_s,
\label{eq:lambdas_supp}\end{equation}
with draws that have $\tilde\Lambda > 5000$ discarded. This produces a prior that is uniform in $\tilde\Lambda$ between 0 and 5000, as shown in Fig.~\ref{fig:lambda_priors} for all of our three mass priors discussed in the main text. For comparison, we also show the prior on $\tilde\Lambda$ computed assuming independent $\Lambda_{1,2} \sim U[0,3000]$ and the component mass prior $m_{1,2}\sim U[1,2]\,M_\odot$. It can be seen that this prior vanishes as $\tilde\Lambda \rightarrow 0$ and so can bias the posterior at low values of   $\tilde\Lambda$. In addition to the physical requirement of a common EOS constraint, the prior used in the common EOS analysis is uniform as $\tilde\Lambda \rightarrow 0$, allowing us to fully explore likelihoods in this region, and set lower bounds on our credible intervals.

\textit{Results}---Fig.~\ref{fig:posterior_overlap} shows the posterior probability densities for the parameters of interest in our study: the source frame chirp mass $\mathcal{M}^\mathrm{src}$; the mass ratio $q=m_2/m_1$; the source frame component masses $m_{1,2}^\mathrm{src}$ (which are functions of $\mathcal{M}^\mathrm{src}$ and $q$); the effective spin $\chi_\mathrm{eff} = (m_1 \chi_1 + m_2 \chi_2) / (m_1 + m_2)$; and the binary tidal deformability $\tilde{\Lambda}$. Posterior probability densities are shown for the uniform mass prior, double neutron star mass prior, and the Galactic neutron star mass prior analyses with 20~Hz low-frequency cutoff, and the uniform mass prior analyses with 25~Hz low-frequency cutoff. All the four analyses had the common EOS constraint and the causal $\Lambda(m)$ lower limit imposed. Electronic files containing the thinned posterior probability densities and an IPython notebook \cite{PER-GRA:2007} for manipulating these data are available at Ref.~\cite{gw170817commoneos}.

\begin{figure*}
  \includegraphics[width=\textwidth]{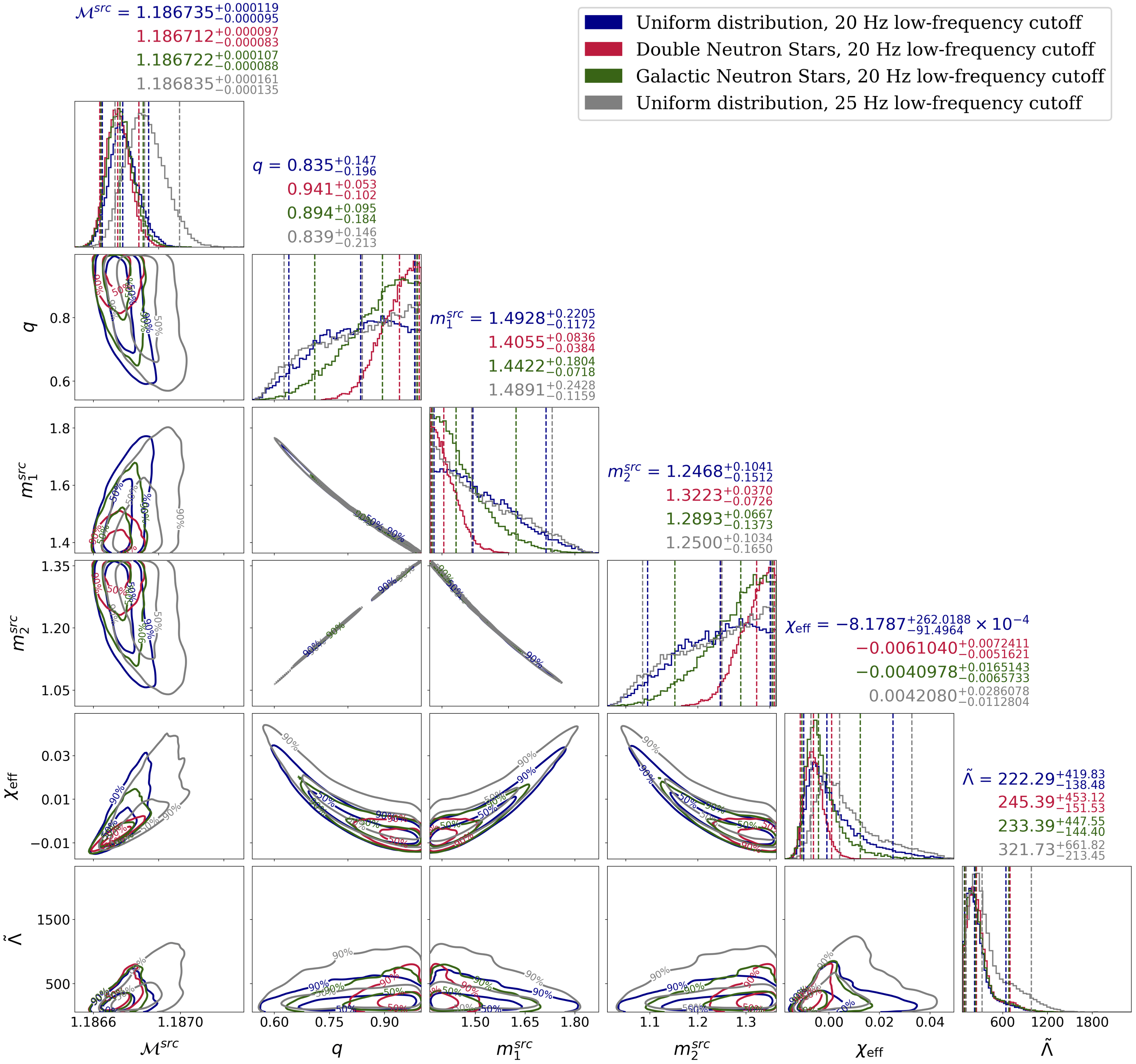}
  \caption{Posterior distributions for the source frame chirp mass $\mathcal{M}^{\rm src}$, mass ratio $q$, source frame primary mass $m_1^{\rm src}$ and secondary mass $m_2^{\rm src}$, effective spin $\chi_{\rm eff}$, and binary deformability parameter $\tilde{\Lambda}$ from parameter estimation analyses with three different choices of mass priors. The posteriors represented in blue are from the analysis using a uniform prior on component masses, $m_{1,2} \sim U[1,2]\, M_\odot$, and 20~Hz low-frequency cutoff. The posteriors represented in red are from the analysis using a Gaussian mass prior for component masses $m_{1,2} \sim N(\mu = 1.33, \sigma = 0.09)\, M_\odot$ known from radio observations of neutron stars in double neutron star (DNS) systems, and 20~Hz low-frequency cutoff. The posteriors represented in green are from the analysis using the observed mass distributions of recycled and slow pulsars in the Galaxy with $m_1 \sim N(\mu = 1.54, \sigma = 0.23)\, M_\odot$ and $m_2 \sim N(\mu = 1.49, \sigma = 0.19)\, M_\odot$~\cite{Ozel:2016oaf}, and 20~Hz low-frequency cutoff. The posteriors represented in gray are from the analysis using a uniform prior on component masses, $m_{1,2} \sim U[1,2]\, M_\odot$, and 25~Hz low-frequency cutoff. All four analyses had the common EOS constraint and the causal $\Lambda(m)$ lower limit imposed. The one-dimensional plots show marginalized probability density functions for the parameters. The dashed lines on the one-dimensional histograms represent the 5$\%$, 50$\%$ and 95$\%$ percentiles for each analysis, the values of which are quoted in the titles of the histograms. The 2D plots show 50$\%$ and 90$\%$ credible regions for the different pairs of parameters. Comparison between the analyses with low-frequency cutoff 20~Hz (blue) and 25~Hz (gray) for the uniform mass prior case shows that extending from 25~Hz to 20~Hz better constrains $\mathcal{M}$, which improves the measurement of $\tilde\Lambda$ by eliminating a region of the posterior with higher values $\mathcal{M}$ and high $\tilde\Lambda$.
  \label{fig:posterior_overlap}}
\end{figure*}

%

%% file: erratum.tex
In the Letter~\cite{De:2018}, we claimed that the GW170817 gravitational-wave data strongly favors models with the common equation of state assumption over those without this assumption. Our statement was based on the values of the Bayes factors that we obtained on comparing the evidences from our parameter estimation analysis of the data using the model with the common equation of state constraint against the model without the constraint. In this erratum, we retract the statement mentioned above, and correct the Bayes factor values reported in the Letter.

Our parameter estimation analysis was performed using the PyCBC Inference software~\cite{Biwer:2018osg} with the parallel-tempered EMCEE sampler~\cite{emcee,vousden:2016,mcmc}, that samples the parameter space using an ensemble of Markov chains at different temperatures. The chains at each temperature sample from a posterior modified by the inverse temperature given by
\begin{equation}
p(\vec{\theta}|\vec{d}(t))_T = p(\vec{\theta}) p(\vec{d}(t)|\vec{\theta})^{1/T}.
\label{eq:postpdf}
\end{equation}
The chains in the colder temperatures ($T \rightarrow 0$) are efficient at finding the peaks of the likelihood ($\mathcal{L} = p(\vec{d}(t)|\vec{\theta})$), and they sample from the posterior. Therefore, samples from these chains are used in the measurement of parameters of of the signal in the data. The chains in the hotter temperatures ($T \rightarrow \infty$) are used in exploring more of the parameter space, they sample from the prior, and help the colder chains find the peak of the likelihood. 

The evidence in a particular EMCEE run is computed using the thermodynamic integration method which integrates the average logarithm of the likelihood $\langle \ln \mathcal{L} \rangle$ as a function of the inverse temperature $\beta$, employing the trapezium rule. Using a sufficiently large number of temperatures placed at the correct locations along the $\langle \ln \mathcal{L} \rangle - \beta$ curve is an important criteria for an accurate measurement of the evidence. It has been pointed out in previous work, that it is important to have a high density of $\beta$ points in locations along the curve where $\langle \ln \mathcal{L} \rangle$ changes substantially with $\beta$ \cite{Liu:2016}. While these strategies are not necessary in helping the chains in the coldest temperature contribute to an accurate measurement of the parameters of the signal, these are indeed important for an accurate computation of the evidence supporting the model used to match the data. When comparing between two models, errors in the measurement of the evidence get propagated to the calculation of the Bayes factor, causing a significant drift from its true value.

\begin{figure}[t]
  \includegraphics[width=\columnwidth]{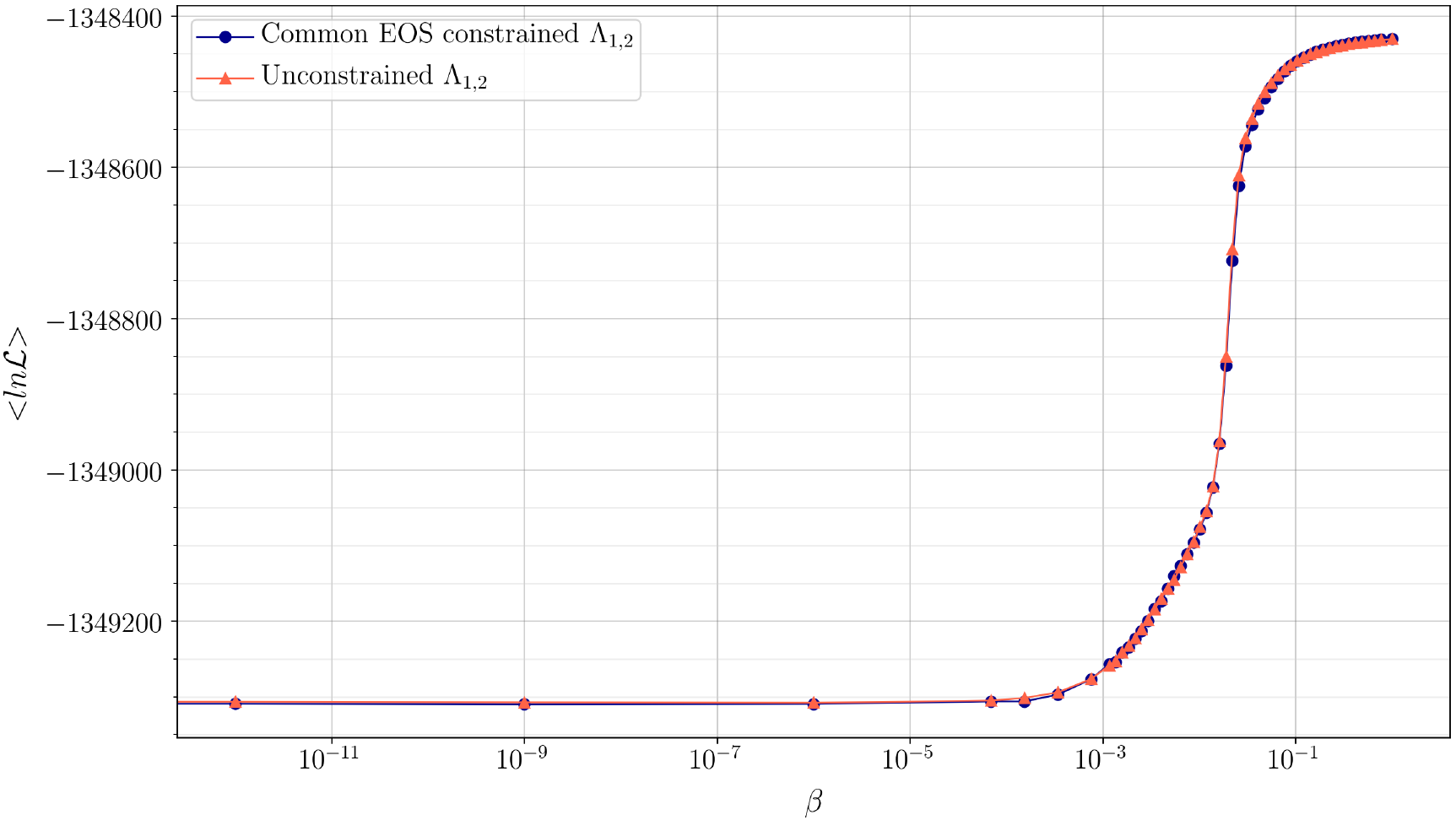}
  \caption{The average logarithm of the likelihood as a function of inverse temperature from parameter estimation analyses of the GW170817 gravitational-wave data using the common equation of state constrained model (in blue) and unconstrained model (in orange) with the uniform mass prior distribution described in the Letter~\cite{De:2018}. The evidence for a model is obtained by integrating its $\ln \mathcal{L}(\beta)$ function. The ratio of the evidences from the models gives the bayes factor $\mathcal{B}$ indicating which model is favored by the data. It can be seen directly from the plot that the $\ln \mathcal{L}(\beta)$ curves for the two models are very similar to each other, resulting in similar values of evidences, and a $\mathcal{B}$ of order unity. This indicates that the gravitational-wave data does not show a substantial preference between the common equation of state constrained and unconstrained models of the analyses in~\cite{De:2018} based on the values of the evidences for the two models.  
\label{fig:lnl_beta}%
\vspace*{-0.5cm}%
}
\end{figure}

We had not taken into account some of these facts in our computation of the evidences for the models in the Letter~\cite{De:2018}. Pursuing our study in \cite{De:2018} further, we ran our analyses with an increased number of temperatures, incorporating the strategies mentioned above to correct our measurements of the evidences using the thermodynamic integration method. Comparing between our model assuming the common equation of state constraint and the model without that constraint for the three mass prior cases, we now obtain Bayes factors of the order unity, instead of 369, 125, and 612 as reported in the Letter. Fig.~\ref{fig:lnl_beta} shows a comparison of the $\langle \ln \mathcal{L} \rangle - \beta$ curves from our common equation of state constrained and unconstrained analyses for the uniform mass prior case. Based on these results, we state that the gravitational-wave data does not indicate a clear preference between the common equation of state constrained and unconstrained models. We will investigate further details such as effects of waveform systematics on the values of the Bayes factors, and explore other methods of calculation of the evidence in a future work. Note that the corrections in this Erratum do not affect any of the parameter measurements, and the primary results of the Letter~\cite{De:2018} remain unchanged.

We thank Collin Capano, Reed Essick, Alexander Nitz, Steven Reyes, and Nevin Weinberg for useful discussions.
%


%% file: main.bbl
\begin{thebibliography}{54}%
\makeatletter
\providecommand \@ifxundefined [1]{%
 \@ifx{#1\undefined}
}%
\providecommand \@ifnum [1]{%
 \ifnum #1\expandafter \@firstoftwo
 \else \expandafter \@secondoftwo
 \fi
}%
\providecommand \@ifx [1]{%
 \ifx #1\expandafter \@firstoftwo
 \else \expandafter \@secondoftwo
 \fi
}%
\providecommand \natexlab [1]{#1}%
\providecommand \enquote  [1]{``#1''}%
\providecommand \bibnamefont  [1]{#1}%
\providecommand \bibfnamefont [1]{#1}%
\providecommand \citenamefont [1]{#1}%
\providecommand \href@noop [0]{\@secondoftwo}%
\providecommand \href [0]{\begingroup \@sanitize@url \@href}%
\providecommand \@href[1]{\@@startlink{#1}\@@href}%
\providecommand \@@href[1]{\endgroup#1\@@endlink}%
\providecommand \@sanitize@url [0]{\catcode `\\12\catcode `\$12\catcode
  `\&12\catcode `\#12\catcode `\^12\catcode `\_12\catcode `\%12\relax}%
\providecommand \@@startlink[1]{}%
\providecommand \@@endlink[0]{}%
\providecommand \url  [0]{\begingroup\@sanitize@url \@url }%
\providecommand \@url [1]{\endgroup\@href {#1}{\urlprefix }}%
\providecommand \urlprefix  [0]{URL }%
\providecommand \Eprint [0]{\href }%
\providecommand \doibase [0]{http://dx.doi.org/}%
\providecommand \selectlanguage [0]{\@gobble}%
\providecommand \bibinfo  [0]{\@secondoftwo}%
\providecommand \bibfield  [0]{\@secondoftwo}%
\providecommand \translation [1]{[#1]}%
\providecommand \BibitemOpen [0]{}%
\providecommand \bibitemStop [0]{}%
\providecommand \bibitemNoStop [0]{.\EOS\space}%
\providecommand \EOS [0]{\spacefactor3000\relax}%
\providecommand \BibitemShut  [1]{\csname bibitem#1\endcsname}%
\let\auto@bib@innerbib\@empty
\bibitem [{\citenamefont {Abbott}\ \emph {et~al.}(2017)\citenamefont {Abbott}
  \emph {et~al.}}]{TheLIGOScientific:2017qsa}%
  \BibitemOpen
  \bibfield  {author} {\bibinfo {author} {\bibfnamefont {B.}~\bibnamefont
  {Abbott}} \emph {et~al.},\ }\href {\doibase 10.1103/PhysRevLett.119.161101}
  {\bibfield  {journal} {\bibinfo  {journal} {Phys. Rev. Lett.}\ }\textbf
  {\bibinfo {volume} {119}},\ \bibinfo {pages} {161101} (\bibinfo {year}
  {2017})}\BibitemShut {NoStop}%
\bibitem [{\citenamefont {Thorne}(1987)}]{thorne.k:1987}%
  \BibitemOpen
  \bibfield  {author} {\bibinfo {author} {\bibfnamefont {K.~S.}\ \bibnamefont
  {Thorne}},\ }in\ \href@noop {} {\emph {\bibinfo {booktitle} {Three hundred
  years of gravitation}}},\ \bibinfo {editor} {edited by\ \bibinfo {editor}
  {\bibfnamefont {S.~W.}\ \bibnamefont {Hawking}}\ and\ \bibinfo {editor}
  {\bibfnamefont {W.}~\bibnamefont {Israel}}}\ (\bibinfo  {publisher}
  {Cambridge University Press},\ \bibinfo {address} {Cambridge},\ \bibinfo
  {year} {1987})\ Chap.~\bibinfo {chapter} {9}, pp.\ \bibinfo {pages}
  {330--458}\BibitemShut {NoStop}%
\bibitem [{\citenamefont {Read}\ \emph {et~al.}(2009)\citenamefont {Read},
  \citenamefont {Markakis}, \citenamefont {Shibata}, \citenamefont {Uryu},
  \citenamefont {Creighton},\ and\ \citenamefont {Friedman}}]{Read:2009yp}%
  \BibitemOpen
  \bibfield  {author} {\bibinfo {author} {\bibfnamefont {J.~S.}\ \bibnamefont
  {Read}}, \bibinfo {author} {\bibfnamefont {C.}~\bibnamefont {Markakis}},
  \bibinfo {author} {\bibfnamefont {M.}~\bibnamefont {Shibata}}, \bibinfo
  {author} {\bibfnamefont {K.}~\bibnamefont {Uryu}}, \bibinfo {author}
  {\bibfnamefont {J.~D.~E.}\ \bibnamefont {Creighton}}, \ and\ \bibinfo
  {author} {\bibfnamefont {J.~L.}\ \bibnamefont {Friedman}},\ }\href {\doibase
  10.1103/PhysRevD.79.124033} {\bibfield  {journal} {\bibinfo  {journal} {Phys.
  Rev.}\ }\textbf {\bibinfo {volume} {D79}},\ \bibinfo {pages} {124033}
  (\bibinfo {year} {2009})}\BibitemShut {NoStop}%
\bibitem [{\citenamefont {Flanagan}\ and\ \citenamefont
  {Hinderer}(2008)}]{Flanagan:2007ix}%
  \BibitemOpen
  \bibfield  {author} {\bibinfo {author} {\bibfnamefont {\'{E}.~\'{E}.}\ \bibnamefont
  {Flanagan}}\ and\ \bibinfo {author} {\bibfnamefont {T.}~\bibnamefont
  {Hinderer}},\ }\href {\doibase 10.1103/PhysRevD.77.021502} {\bibfield
  {journal} {\bibinfo  {journal} {Phys. Rev.}\ }\textbf {\bibinfo {volume}
  {D77}},\ \bibinfo {pages} {021502} (\bibinfo {year} {2008})}\BibitemShut
  {NoStop}%
\bibitem [{\citenamefont {Hinderer}(2008)}]{Hinderer:2007mb}%
  \BibitemOpen
  \bibfield  {author} {\bibinfo {author} {\bibfnamefont {T.}~\bibnamefont
  {Hinderer}},\ }\href {\doibase 10.1086/533487} {\bibfield  {journal}
  {\bibinfo  {journal} {Astrophys. J.}\ }\textbf {\bibinfo {volume} {677}},\
  \bibinfo {pages} {1216} (\bibinfo {year} {2008})}\BibitemShut {NoStop}%
\bibitem [{\citenamefont {Gralla}(2018)}]{Gralla:2017djj}%
  \BibitemOpen
  \bibfield  {author} {\bibinfo {author} {\bibfnamefont {S.~E.}\ \bibnamefont
  {Gralla}},\ }\href {\doibase 10.1088/1361-6382/aab186} {\bibfield  {journal}
  {\bibinfo  {journal} {Class. Quant. Grav.}\ }\textbf {\bibinfo {volume}
  {35}},\ \bibinfo {pages} {085002} (\bibinfo {year} {2018})}\BibitemShut
  {NoStop}%
\bibitem [{\citenamefont {Biwer}\ \emph {et~al.}(2018)\citenamefont {Biwer},
  \citenamefont {Capano}, \citenamefont {De}, \citenamefont {Cabero},
  \citenamefont {Brown}, \citenamefont {Nitz},\ and\ \citenamefont
  {Raymond}}]{Biwer:2018osg}%
  \BibitemOpen
  \bibfield  {author} {\bibinfo {author} {\bibfnamefont {C.~M.}\ \bibnamefont
  {Biwer}}, \bibinfo {author} {\bibfnamefont {C.~D.}\ \bibnamefont {Capano}},
  \bibinfo {author} {\bibfnamefont {S.}~\bibnamefont {De}}, \bibinfo {author}
  {\bibfnamefont {M.}~\bibnamefont {Cabero}}, \bibinfo {author} {\bibfnamefont
  {D.~A.}\ \bibnamefont {Brown}}, \bibinfo {author} {\bibfnamefont {A.~H.}\
  \bibnamefont {Nitz}}, \ and\ \bibinfo {author} {\bibfnamefont
  {V.}~\bibnamefont {Raymond}},\ }\href@noop {} {\  (\bibinfo {year} {2018})},\
  \Eprint {http://arxiv.org/abs/1807.10312} {arXiv:1807.10312 [astro-ph.IM]}
  \BibitemShut {NoStop}%
\bibitem [{\citenamefont {Nitz}\ \emph {et~al.}(2018)\citenamefont {Nitz} \emph
  {et~al.}}]{alex_nitz_2018_1208115}%
  \BibitemOpen
  \bibfield  {author} {\bibinfo {author} {\bibfnamefont {A.}~\bibnamefont
  {Nitz}} \emph {et~al.},\ }\href {\doibase 10.5281/zenodo.1208115} {\emph
  {\bibinfo {title} {PyCBC v1.9.4}}} (\bibinfo {year} {2018})\BibitemShut
  {NoStop}%
\bibitem [{\citenamefont {Foreman-Mackey}\ \emph {et~al.}(2013)\citenamefont
  {Foreman-Mackey}, \citenamefont {Hogg}, \citenamefont {Lang},\ and\
  \citenamefont {Goodman}}]{emcee}%
  \BibitemOpen
  \bibfield  {author} {\bibinfo {author} {\bibfnamefont {D.}~\bibnamefont
  {Foreman-Mackey}}, \bibinfo {author} {\bibfnamefont {D.~W.}\ \bibnamefont
  {Hogg}}, \bibinfo {author} {\bibfnamefont {D.}~\bibnamefont {Lang}}, \ and\
  \bibinfo {author} {\bibfnamefont {J.}~\bibnamefont {Goodman}},\ }\href
  {http://stacks.iop.org/1538-3873/125/i=925/a=306} {\bibfield  {journal}
  {\bibinfo  {journal} {Publications of the Astronomical Society of the
  Pacific}\ }\textbf {\bibinfo {volume} {125}},\ \bibinfo {pages} {306}
  (\bibinfo {year} {2013})}\BibitemShut {NoStop}%
\bibitem [{\citenamefont {Soares-Santos}\ \emph {et~al.}(2017)\citenamefont
  {Soares-Santos} \emph {et~al.}}]{Soares-Santos:2017lru}%
  \BibitemOpen
  \bibfield  {author} {\bibinfo {author} {\bibfnamefont {M.}~\bibnamefont
  {Soares-Santos}} \emph {et~al.},\ }\href {\doibase 10.3847/2041-8213/aa9059}
  {\bibfield  {journal} {\bibinfo  {journal} {Astrophys. J.}\ }\textbf
  {\bibinfo {volume} {848}},\ \bibinfo {pages} {L16} (\bibinfo {year}
  {2017})}\BibitemShut {NoStop}%
\bibitem [{\citenamefont {Cantiello}\ \emph {et~al.}(2018)\citenamefont
  {Cantiello} \emph {et~al.}}]{Cantiello:2018ffy}%
  \BibitemOpen
  \bibfield  {author} {\bibinfo {author} {\bibfnamefont {M.}~\bibnamefont
  {Cantiello}} \emph {et~al.},\ }\href {\doibase 10.3847/2041-8213/aaad64}
  {\bibfield  {journal} {\bibinfo  {journal} {Astrophys. J.}\ }\textbf
  {\bibinfo {volume} {854}},\ \bibinfo {pages} {L31} (\bibinfo {year}
  {2018})}\BibitemShut {NoStop}%
\bibitem [{\citenamefont {\"{O}zel}\ and\ \citenamefont
  {Freire}(2016)}]{Ozel:2016oaf}%
  \BibitemOpen
  \bibfield  {author} {\bibinfo {author} {\bibfnamefont {F.}~\bibnamefont
  {\"{O}zel}}\ and\ \bibinfo {author} {\bibfnamefont {P.}~\bibnamefont
  {Freire}},\ }\href {\doibase 10.1146/annurev-astro-081915-023322} {\bibfield
  {journal} {\bibinfo  {journal} {Ann. Rev. Astron. Astrophys.}\ }\textbf
  {\bibinfo {volume} {54}},\ \bibinfo {pages} {401} (\bibinfo {year}
  {2016})}\BibitemShut {NoStop}%
\bibitem [{\citenamefont {Lattimer}\ and\ \citenamefont
  {Prakash}(2016)}]{Lattimer:2015nhk}%
  \BibitemOpen
  \bibfield  {author} {\bibinfo {author} {\bibfnamefont {J.~M.}\ \bibnamefont
  {Lattimer}}\ and\ \bibinfo {author} {\bibfnamefont {M.}~\bibnamefont
  {Prakash}},\ }\href {\doibase 10.1016/j.physrep.2015.12.005} {\bibfield
  {journal} {\bibinfo  {journal} {Phys. Rept.}\ }\textbf {\bibinfo {volume}
  {621}},\ \bibinfo {pages} {127} (\bibinfo {year} {2016})}\BibitemShut
  {NoStop}%
\bibitem [{\citenamefont {Lattimer}(2012)}]{Lattimer:2012nd}%
  \BibitemOpen
  \bibfield  {author} {\bibinfo {author} {\bibfnamefont {J.~M.}\ \bibnamefont
  {Lattimer}},\ }\href {\doibase 10.1146/annurev-nucl-102711-095018} {\bibfield
   {journal} {\bibinfo  {journal} {Ann. Rev. Nucl. Part. Sci.}\ }\textbf
  {\bibinfo {volume} {62}},\ \bibinfo {pages} {485} (\bibinfo {year}
  {2012})}\BibitemShut {NoStop}%
\bibitem [{\citenamefont {Antoniadis}\ \emph {et~al.}(2013)\citenamefont
  {Antoniadis} \emph {et~al.}}]{Antoniadis:2013pzd}%
  \BibitemOpen
  \bibfield  {author} {\bibinfo {author} {\bibfnamefont {J.}~\bibnamefont
  {Antoniadis}} \emph {et~al.},\ }\href {\doibase 10.1126/science.1233232}
  {\bibfield  {journal} {\bibinfo  {journal} {Science}\ }\textbf {\bibinfo
  {volume} {340}},\ \bibinfo {pages} {1233232} (\bibinfo {year}
  {2013})}\BibitemShut {NoStop}%
\bibitem [{\citenamefont {Oppenheimer}\ and\ \citenamefont
  {Volkoff}(1939)}]{Oppenheimer:1939ne}%
  \BibitemOpen
  \bibfield  {author} {\bibinfo {author} {\bibfnamefont {J.~R.}\ \bibnamefont
  {Oppenheimer}}\ and\ \bibinfo {author} {\bibfnamefont {G.~M.}\ \bibnamefont
  {Volkoff}},\ }\href {\doibase 10.1103/PhysRev.55.374} {\bibfield  {journal}
  {\bibinfo  {journal} {Phys. Rev.}\ }\textbf {\bibinfo {volume} {55}},\
  \bibinfo {pages} {374} (\bibinfo {year} {1939})}\BibitemShut {NoStop}%
\bibitem [{\citenamefont {Postnikov}\ \emph {et~al.}(2010)\citenamefont
  {Postnikov}, \citenamefont {Prakash},\ and\ \citenamefont
  {Lattimer}}]{Postnikov:2010yn}%
  \BibitemOpen
  \bibfield  {author} {\bibinfo {author} {\bibfnamefont {S.}~\bibnamefont
  {Postnikov}}, \bibinfo {author} {\bibfnamefont {M.}~\bibnamefont {Prakash}},
  \ and\ \bibinfo {author} {\bibfnamefont {J.~M.}\ \bibnamefont {Lattimer}},\
  }\href {\doibase 10.1103/PhysRevD.82.024016} {\bibfield  {journal} {\bibinfo
  {journal} {Phys. Rev.}\ }\textbf {\bibinfo {volume} {D82}},\ \bibinfo {pages}
  {024016} (\bibinfo {year} {2010})}\BibitemShut {NoStop}%
\bibitem{supp}
See Supplemental Material for descriptions supporting our implementation of the common radius and causal constraints, technical details of the parameter estimation analysis, full posterior distributions from the common EOS runs, and comparison of results to Ref~\cite{TheLIGOScientific:2017qsa}. The Supplemental Material includes
Refs. [1,7-14,16,19-39].%
\bibitem [{\citenamefont {Driggers}\ \emph {et~al.}(2017)\citenamefont
  {Driggers}, \citenamefont {Vitale}, \citenamefont {Lundgren}, \citenamefont
  {Evans}, \citenamefont {Kawabe}, \citenamefont {Dwyer}, \citenamefont
  {Izumi},\ and\ \citenamefont {Fritschel}}]{gw170817-noise}%
  \BibitemOpen
  \bibfield  {author} {\bibinfo {author} {\bibfnamefont {J.}~\bibnamefont
  {Driggers}}, \bibinfo {author} {\bibfnamefont {S.}~\bibnamefont {Vitale}},
  \bibinfo {author} {\bibfnamefont {A.}~\bibnamefont {Lundgren}}, \bibinfo
  {author} {\bibfnamefont {M.}~\bibnamefont {Evans}}, \bibinfo {author}
  {\bibfnamefont {K.}~\bibnamefont {Kawabe}}, \bibinfo {author} {\bibfnamefont
  {S.}~\bibnamefont {Dwyer}}, \bibinfo {author} {\bibfnamefont
  {K.}~\bibnamefont {Izumi}}, \ and\ \bibinfo {author} {\bibfnamefont
  {P.}~\bibnamefont {Fritschel}},\ }\href@noop {} {\enquote {\bibinfo {title}
  {Offline noise subtraction for {Advanced LIGO}},}\ } (\bibinfo {year}
  {2017}),\ \bibinfo {note}
  {{https://dcc.ligo.org/LIGO-P1700260/public}}\BibitemShut {NoStop}%
\bibitem [{\citenamefont {Welch}(1967)}]{1161901}%
  \BibitemOpen
  \bibfield  {author} {\bibinfo {author} {\bibfnamefont {P.}~\bibnamefont
  {Welch}},\ }\href {\doibase 10.1109/TAU.1967.1161901} {\bibfield  {journal}
  {\bibinfo  {journal} {IEEE Transactions on Audio and Electroacoustics}\
  }\textbf {\bibinfo {volume} {15}},\ \bibinfo {pages} {70} (\bibinfo {year}
  {1967})}\BibitemShut {NoStop}%
\bibitem [{\citenamefont {Allen}\ \emph {et~al.}(2012)\citenamefont {Allen},
  \citenamefont {Anderson}, \citenamefont {Brady}, \citenamefont {Brown},\ and\
  \citenamefont {Creighton}}]{Allen:2005fk}%
  \BibitemOpen
  \bibfield  {author} {\bibinfo {author} {\bibfnamefont {B.}~\bibnamefont
  {Allen}}, \bibinfo {author} {\bibfnamefont {W.~G.}\ \bibnamefont {Anderson}},
  \bibinfo {author} {\bibfnamefont {P.~R.}\ \bibnamefont {Brady}}, \bibinfo
  {author} {\bibfnamefont {D.~A.}\ \bibnamefont {Brown}}, \ and\ \bibinfo
  {author} {\bibfnamefont {J.~D.~E.}\ \bibnamefont {Creighton}},\ }\href
  {\doibase 10.1103/PhysRevD.85.122006} {\bibfield  {journal} {\bibinfo
  {journal} {Phys. Rev.}\ }\textbf {\bibinfo {volume} {D85}},\ \bibinfo {pages}
  {122006} (\bibinfo {year} {2012})}\BibitemShut {NoStop}%
\bibitem [{\citenamefont {Damour}\ \emph {et~al.}(2012)\citenamefont {Damour},
  \citenamefont {Nagar},\ and\ \citenamefont {Villain}}]{Damour:2012yf}%
  \BibitemOpen
  \bibfield  {author} {\bibinfo {author} {\bibfnamefont {T.}~\bibnamefont
  {Damour}}, \bibinfo {author} {\bibfnamefont {A.}~\bibnamefont {Nagar}}, \
  and\ \bibinfo {author} {\bibfnamefont {L.}~\bibnamefont {Villain}},\ }\href
  {\doibase 10.1103/PhysRevD.85.123007} {\bibfield  {journal} {\bibinfo
  {journal} {Phys. Rev.}\ }\textbf {\bibinfo {volume} {D85}},\ \bibinfo {pages}
  {123007} (\bibinfo {year} {2012})}\BibitemShut {NoStop}%
\bibitem [{\citenamefont {Littenberg}\ and\ \citenamefont
  {Cornish}(2015)}]{Littenberg:2014oda}%
  \BibitemOpen
  \bibfield  {author} {\bibinfo {author} {\bibfnamefont {T.~B.}\ \bibnamefont
  {Littenberg}}\ and\ \bibinfo {author} {\bibfnamefont {N.~J.}\ \bibnamefont
  {Cornish}},\ }\href {\doibase 10.1103/PhysRevD.91.084034} {\bibfield
  {journal} {\bibinfo  {journal} {Phys. Rev.}\ }\textbf {\bibinfo {volume}
  {D91}},\ \bibinfo {pages} {084034} (\bibinfo {year} {2015})}\BibitemShut
  {NoStop}%
\bibitem [{\citenamefont {Mercer}\ \emph {et~al.}(2017)\citenamefont {Mercer}
  \emph {et~al.}}]{lal}%
  \BibitemOpen
  \bibfield  {author} {\bibinfo {author} {\bibfnamefont {R.~A.}\ \bibnamefont
  {Mercer}} \emph {et~al.},\ }\href@noop {} {\enquote {\bibinfo {title} {{LIGO
  Algorithm Library}},}\ } (\bibinfo {year} {2017}),\ \bibinfo {note}
  {{https://git.ligo.org/lscsoft/lalsuite}}\BibitemShut {NoStop}%
\bibitem [{\citenamefont {P\'erez}\ and\ \citenamefont
  {Granger}(2007)}]{PER-GRA:2007}%
  \BibitemOpen
  \bibfield  {author} {\bibinfo {author} {\bibfnamefont {F.}~\bibnamefont
  {P\'erez}}\ and\ \bibinfo {author} {\bibfnamefont {B.~E.}\ \bibnamefont
  {Granger}},\ }\href {\doibase 10.1109/MCSE.2007.53} {\bibfield  {journal}
  {\bibinfo  {journal} {Computing in Science and Engineering}\ }\textbf
  {\bibinfo {volume} {9}},\ \bibinfo {pages} {21} (\bibinfo {year}
  {2007})}\BibitemShut {NoStop}%
\bibitem [{\citenamefont {Vallisneri}\ \emph {et~al.}(2015)\citenamefont
  {Vallisneri}, \citenamefont {Kanner}, \citenamefont {Williams}, \citenamefont
  {Weinstein},\ and\ \citenamefont {Stephens}}]{Vallisneri:2014vxa}%
  \BibitemOpen
  \bibfield  {author} {\bibinfo {author} {\bibfnamefont {M.}~\bibnamefont
  {Vallisneri}}, \bibinfo {author} {\bibfnamefont {J.}~\bibnamefont {Kanner}},
  \bibinfo {author} {\bibfnamefont {R.}~\bibnamefont {Williams}}, \bibinfo
  {author} {\bibfnamefont {A.}~\bibnamefont {Weinstein}}, \ and\ \bibinfo
  {author} {\bibfnamefont {B.}~\bibnamefont {Stephens}},\ }\href
  {http://stacks.iop.org/1742-6596/610/i=1/a=012021} {\bibfield  {journal}
  {\bibinfo  {journal} {Journal of Physics: Conference Series}\ }\textbf
  {\bibinfo {volume} {610}},\ \bibinfo {pages} {012021} (\bibinfo {year}
  {2015})}\BibitemShut {NoStop}%
\bibitem [{\citenamefont {Blackburn}\ \emph {et~al.}(2017)\citenamefont
  {Blackburn} \emph {et~al.}}]{gw170817-losc}%
  \BibitemOpen
  \bibfield  {author} {\bibinfo {author} {\bibfnamefont {K.}~\bibnamefont
  {Blackburn}} \emph {et~al.},\ }\href {\doibase doi:10.7935/K5B8566F} {\emph
  {\bibinfo {title} {{LOSC CLN Data Products for GW170817}}}} (\bibinfo {year}
  {2017})\BibitemShut {NoStop}%
\bibitem [{\citenamefont {Vousden}\ \emph {et~al.}(2016)\citenamefont
  {Vousden}, \citenamefont {Farr},\ and\ \citenamefont
  {Mandel}}]{vousden:2016}%
  \BibitemOpen
  \bibfield  {author} {\bibinfo {author} {\bibfnamefont {W.~D.}\ \bibnamefont
  {Vousden}}, \bibinfo {author} {\bibfnamefont {W.~M.}\ \bibnamefont {Farr}}, \
  and\ \bibinfo {author} {\bibfnamefont {I.}~\bibnamefont {Mandel}},\ }\href
  {\doibase 10.1093/mnras/stv2422} {\bibfield  {journal} {\bibinfo  {journal}
  {Monthly Notices of the Royal Astronomical Society}\ }\textbf {\bibinfo
  {volume} {455}},\ \bibinfo {pages} {1919} (\bibinfo {year}
  {2016})}\BibitemShut {NoStop}%
\bibitem [{\citenamefont {Goodman~J.}(2010)}]{mcmc}%
  \BibitemOpen
  \bibfield  {author} {\bibinfo {author} {\bibfnamefont {W.~J.}\ \bibnamefont
  {Goodman~J.}},\ }\href {\doibase 10.2140/camcos.2010.5.65} {\bibfield
  {journal} {\bibinfo  {journal} {Commun. Appl. Math. Comput. Sci.}\ }\textbf
  {\bibinfo {volume} {5}},\ \bibinfo {pages} {65} (\bibinfo {year}
  {2010})}\BibitemShut {NoStop}%
\bibitem [{\citenamefont {Buonanno}\ \emph {et~al.}(2009)\citenamefont
  {Buonanno}, \citenamefont {Iyer}, \citenamefont {Ochsner}, \citenamefont
  {Pan},\ and\ \citenamefont {Sathyaprakash}}]{Buonanno:2009zt}%
  \BibitemOpen
  \bibfield  {author} {\bibinfo {author} {\bibfnamefont {A.}~\bibnamefont
  {Buonanno}}, \bibinfo {author} {\bibfnamefont {B.~R.}\ \bibnamefont {Iyer}},
  \bibinfo {author} {\bibfnamefont {E.}~\bibnamefont {Ochsner}}, \bibinfo
  {author} {\bibfnamefont {Y.}~\bibnamefont {Pan}}, \ and\ \bibinfo {author}
  {\bibfnamefont {B.~S.}\ \bibnamefont {Sathyaprakash}},\ }\href {\doibase
  10.1103/PhysRevD.80.084043} {\bibfield  {journal} {\bibinfo  {journal} {Phys.
  Rev. D}\ }\textbf {\bibinfo {volume} {80}},\ \bibinfo {pages} {084043}
  (\bibinfo {year} {2009})}\BibitemShut {NoStop}%
\bibitem [{\citenamefont {Arun}\ \emph {et~al.}(2009)\citenamefont {Arun},
  \citenamefont {Buonanno}, \citenamefont {Faye},\ and\ \citenamefont
  {Ochsner}}]{Arun:2008kb}%
  \BibitemOpen
  \bibfield  {author} {\bibinfo {author} {\bibfnamefont {K.~G.}\ \bibnamefont
  {Arun}}, \bibinfo {author} {\bibfnamefont {A.}~\bibnamefont {Buonanno}},
  \bibinfo {author} {\bibfnamefont {G.}~\bibnamefont {Faye}}, \ and\ \bibinfo
  {author} {\bibfnamefont {E.}~\bibnamefont {Ochsner}},\ }\href {\doibase
  10.1103/PhysRevD.79.104023, 10.1103/PhysRevD.84.049901} {\bibfield  {journal}
  {\bibinfo  {journal} {Phys. Rev.}\ }\textbf {\bibinfo {volume} {D79}},\
  \bibinfo {pages} {104023} (\bibinfo {year} {2009})},\ \bibinfo {note}
  {[Erratum: Phys. Rev.D84,049901(2011)]}\BibitemShut {NoStop}%
\bibitem [{\citenamefont {Mikoczi}\ \emph {et~al.}(2005)\citenamefont
  {Mikoczi}, \citenamefont {Vasuth},\ and\ \citenamefont
  {Gergely}}]{Mikoczi:2005dn}%
  \BibitemOpen
  \bibfield  {author} {\bibinfo {author} {\bibfnamefont {B.}~\bibnamefont
  {Mikoczi}}, \bibinfo {author} {\bibfnamefont {M.}~\bibnamefont {Vasuth}}, \
  and\ \bibinfo {author} {\bibfnamefont {L.~A.}\ \bibnamefont {Gergely}},\
  }\href {\doibase 10.1103/PhysRevD.71.124043} {\bibfield  {journal} {\bibinfo
  {journal} {Phys. Rev.}\ }\textbf {\bibinfo {volume} {D71}},\ \bibinfo {pages}
  {124043} (\bibinfo {year} {2005})}\BibitemShut {NoStop}%
\bibitem [{\citenamefont {Boh\'{e}}\ \emph {et~al.}(2013)\citenamefont
  {Boh\'{e}}, \citenamefont {Marsat},\ and\ \citenamefont
  {Blanchet}}]{Bohe:2013cla}%
  \BibitemOpen
  \bibfield  {author} {\bibinfo {author} {\bibfnamefont {A.}~\bibnamefont
  {Boh\'{e}}}, \bibinfo {author} {\bibfnamefont {S.}~\bibnamefont {Marsat}}, \
  and\ \bibinfo {author} {\bibfnamefont {L.}~\bibnamefont {Blanchet}},\ }\href
  {http://stacks.iop.org/0264-9381/30/i=13/a=135009} {\bibfield  {journal}
  {\bibinfo  {journal} {Classical and Quantum Gravity}\ }\textbf {\bibinfo
  {volume} {30}},\ \bibinfo {pages} {135009} (\bibinfo {year}
  {2013})}\BibitemShut {NoStop}%
\bibitem [{\citenamefont {Vines}\ \emph {et~al.}(2011)\citenamefont {Vines},
  \citenamefont {Flanagan},\ and\ \citenamefont {Hinderer}}]{Vines:2011ud}%
  \BibitemOpen
  \bibfield  {author} {\bibinfo {author} {\bibfnamefont {J.}~\bibnamefont
  {Vines}}, \bibinfo {author} {\bibfnamefont {\'{E}.~\'{E}.}\ \bibnamefont {Flanagan}},
  \ and\ \bibinfo {author} {\bibfnamefont {T.}~\bibnamefont {Hinderer}},\
  }\href {\doibase 10.1103/PhysRevD.83.084051} {\bibfield  {journal} {\bibinfo
  {journal} {Phys. Rev. D}\ }\textbf {\bibinfo {volume} {83}},\ \bibinfo
  {pages} {084051} (\bibinfo {year} {2011})}\BibitemShut {NoStop}%
\bibitem [{\citenamefont {Brown}\ \emph {et~al.}(2012)\citenamefont {Brown},
  \citenamefont {Harry}, \citenamefont {Lundgren},\ and\ \citenamefont
  {Nitz}}]{Brown:2012qf}%
  \BibitemOpen
  \bibfield  {author} {\bibinfo {author} {\bibfnamefont {D.~A.}\ \bibnamefont
  {Brown}}, \bibinfo {author} {\bibfnamefont {I.}~\bibnamefont {Harry}},
  \bibinfo {author} {\bibfnamefont {A.}~\bibnamefont {Lundgren}}, \ and\
  \bibinfo {author} {\bibfnamefont {A.~H.}\ \bibnamefont {Nitz}},\ }\href
  {\doibase 10.1103/PhysRevD.86.084017} {\bibfield  {journal} {\bibinfo
  {journal} {Phys. Rev. D}\ }\textbf {\bibinfo {volume} {86}},\ \bibinfo
  {pages} {084017} (\bibinfo {year} {2012})}\BibitemShut {NoStop}%
\bibitem [{\citenamefont {De}\ \emph {et~al.}(2018)\citenamefont {De} \emph
  {et~al.}}]{gw170817commoneos}%
  \BibitemOpen
  \bibfield  {author} {\bibinfo {author} {\bibfnamefont {S.}~\bibnamefont {De}}
  \emph {et~al.},\ }\href {https://github.com/sugwg/gw170817-common-eos} {\emph
  {\bibinfo {title} {SUGWG GitHub Repository}}} (\bibinfo {year}
  {2018})\BibitemShut {NoStop}%
\bibitem [{\citenamefont {Zhao}\ and\ \citenamefont
  {Lattimer}(2018)}]{Zhao:2018nyf}%
  \BibitemOpen
  \bibfield  {author} {\bibinfo {author} {\bibfnamefont {T.}~\bibnamefont
  {Zhao}}\ and\ \bibinfo {author} {\bibfnamefont {J.~M.}\ \bibnamefont
  {Lattimer}},\ }\href@noop {} {\  (\bibinfo {year} {2018})},\ \Eprint
  {http://arxiv.org/abs/1808.02858} {arXiv:1808.02858 [astro-ph.HE]}
  \BibitemShut {NoStop}%
\bibitem [{\citenamefont {Tauris}\ \emph {et~al.}(2017)\citenamefont {Tauris}
  \emph {et~al.}}]{Tauris:2017omb}%
  \BibitemOpen
  \bibfield  {author} {\bibinfo {author} {\bibfnamefont {T.~M.}\ \bibnamefont
  {Tauris}} \emph {et~al.},\ }\href {\doibase 10.3847/1538-4357/aa7e89}
  {\bibfield  {journal} {\bibinfo  {journal} {Astrophys. J.}\ }\textbf
  {\bibinfo {volume} {846}},\ \bibinfo {pages} {170} (\bibinfo {year}
  {2017})}\BibitemShut {NoStop}%
\bibitem [{\citenamefont {Abbott}\ \emph {et~al.}(2018)\citenamefont {Abbott}
  \emph {et~al.}}]{Abbott:2018wiz}%
  \BibitemOpen
  \bibfield  {author} {\bibinfo {author} {\bibfnamefont {B.~P.}\ \bibnamefont
  {Abbott}} \emph {et~al.} (\bibinfo {collaboration} {Virgo, LIGO
  Scientific}),\ }\href@noop {} {\  (\bibinfo {year} {2018})},\ \Eprint
  {http://arxiv.org/abs/1805.11579} {arXiv:1805.11579 [gr-qc]} \BibitemShut
  {NoStop}%
\bibitem [{\citenamefont {Christensen}\ and\ \citenamefont
  {Meyer}(2001)}]{Christensen:2001cr}%
  \BibitemOpen
  \bibfield  {author} {\bibinfo {author} {\bibfnamefont {N.}~\bibnamefont
  {Christensen}}\ and\ \bibinfo {author} {\bibfnamefont {R.}~\bibnamefont
  {Meyer}},\ }\href {\doibase 10.1103/PhysRevD.64.022001} {\bibfield  {journal}
  {\bibinfo  {journal} {Phys. Rev.}\ }\textbf {\bibinfo {volume} {D64}},\
  \bibinfo {pages} {022001} (\bibinfo {year} {2001})}\BibitemShut {NoStop}%
\bibitem [{\citenamefont {Rover}\ \emph {et~al.}(2007)\citenamefont {Rover},
  \citenamefont {Meyer},\ and\ \citenamefont {Christensen}}]{Rover:2006bb}%
  \BibitemOpen
  \bibfield  {author} {\bibinfo {author} {\bibfnamefont {C.}~\bibnamefont
  {R\"{o}ver}}, \bibinfo {author} {\bibfnamefont {R.}~\bibnamefont {Meyer}}, \ and\
  \bibinfo {author} {\bibfnamefont {N.}~\bibnamefont {Christensen}},\
  }\href {\doibase 10.1103/PhysRevD.75.062004} {\bibfield  {journal} {\bibinfo  {journal} {Phys. Rev.}\
  }\textbf {\bibinfo {volume} {D75}},\ \bibinfo {pages} {062004} (\bibinfo
  {year} {2007})}\BibitemShut {NoStop}%
\bibitem [{\citenamefont {Sathyaprakash}\ and\ \citenamefont
  {Dhurandhar}(1991)}]{Sathyaprakash:1991mt}%
  \BibitemOpen
  \bibfield  {author} {\bibinfo {author} {\bibfnamefont {B.~S.}\ \bibnamefont
  {Sathyaprakash}}\ and\ \bibinfo {author} {\bibfnamefont {S.~V.}\ \bibnamefont
  {Dhurandhar}},\ }\href {\doibase 10.1103/PhysRevD.44.3819} {\bibfield
  {journal} {\bibinfo  {journal} {Phys. Rev.}\ }\textbf {\bibinfo {volume}
  {D44}},\ \bibinfo {pages} {3819} (\bibinfo {year} {1991})}\BibitemShut
  {NoStop}%
\bibitem [{\citenamefont {Agathos}\ \emph {et~al.}(2015)\citenamefont
  {Agathos}, \citenamefont {Meidam}, \citenamefont {Del~Pozzo}, \citenamefont
  {Li}, \citenamefont {Tompitak}, \citenamefont {Veitch}, \citenamefont
  {Vitale},\ and\ \citenamefont {Van Den~Broeck}}]{Agathos:2015uaa}%
  \BibitemOpen
  \bibfield  {author} {\bibinfo {author} {\bibfnamefont {M.}~\bibnamefont
  {Agathos}}, \bibinfo {author} {\bibfnamefont {J.}~\bibnamefont {Meidam}},
  \bibinfo {author} {\bibfnamefont {W.}~\bibnamefont {Del~Pozzo}}, \bibinfo
  {author} {\bibfnamefont {T.~G.~F.}\ \bibnamefont {Li}}, \bibinfo {author}
  {\bibfnamefont {M.}~\bibnamefont {Tompitak}}, \bibinfo {author}
  {\bibfnamefont {J.}~\bibnamefont {Veitch}}, \bibinfo {author} {\bibfnamefont
  {S.}~\bibnamefont {Vitale}}, \ and\ \bibinfo {author} {\bibfnamefont
  {C.}~\bibnamefont {Van Den~Broeck}},\ }\href {\doibase 10.1103/PhysRevD.92.023012} {\bibfield  {journal}
  {\bibinfo  {journal} {Phys. Rev.}\ }\textbf {\bibinfo {volume} {D92}},\
  \bibinfo {pages} {023012} (\bibinfo {year} {2015})}\BibitemShut {NoStop}%
\bibitem [{\citenamefont {Skilling}(2006)}]{skilling2006}%
  \BibitemOpen
  \bibfield  {author} {\bibinfo {author} {\bibfnamefont {J.}~\bibnamefont
  {Skilling}},\ }\href {\doibase 10.1214/06-BA127} {\bibfield  {journal}
  {\bibinfo  {journal} {Bayesian Anal.}\ }\textbf {\bibinfo {volume} {1}},\
  \bibinfo {pages} {833} (\bibinfo {year} {2006})}\BibitemShut {NoStop}%
\bibitem [{\citenamefont {Wade}\ \emph {et~al.}(2014)\citenamefont {Wade},
  \citenamefont {Creighton}, \citenamefont {Ochsner}, \citenamefont {Lackey},
  \citenamefont {Farr}, \citenamefont {Littenberg},\ and\ \citenamefont
  {Raymond}}]{Wade:2014vqa}%
  \BibitemOpen
  \bibfield  {author} {\bibinfo {author} {\bibfnamefont {L.}~\bibnamefont
  {Wade}}, \bibinfo {author} {\bibfnamefont {J.~D.~E.}\ \bibnamefont
  {Creighton}}, \bibinfo {author} {\bibfnamefont {E.}~\bibnamefont {Ochsner}},
  \bibinfo {author} {\bibfnamefont {B.~D.}\ \bibnamefont {Lackey}}, \bibinfo
  {author} {\bibfnamefont {B.~F.}\ \bibnamefont {Farr}}, \bibinfo {author}
  {\bibfnamefont {T.~B.}\ \bibnamefont {Littenberg}}, \ and\ \bibinfo {author}
  {\bibfnamefont {V.}~\bibnamefont {Raymond}},\ }\href {\doibase
  10.1103/PhysRevD.89.103012} {\bibfield  {journal} {\bibinfo  {journal} {Phys.
  Rev.}\ }\textbf {\bibinfo {volume} {D89}},\ \bibinfo {pages} {103012}
  (\bibinfo {year} {2014})}\BibitemShut {NoStop}%
\bibitem [{\citenamefont {Lackey}\ and\ \citenamefont
  {Wade}(2015)}]{Lackey:2014fwa}%
  \BibitemOpen
  \bibfield  {author} {\bibinfo {author} {\bibfnamefont {B.~D.}\ \bibnamefont
  {Lackey}}\ and\ \bibinfo {author} {\bibfnamefont {L.}~\bibnamefont {Wade}},\
  }\href {\doibase 10.1103/PhysRevD.91.043002} {\bibfield  {journal} {\bibinfo
  {journal} {Phys. Rev.}\ }\textbf {\bibinfo {volume} {D91}},\ \bibinfo {pages}
  {043002} (\bibinfo {year} {2015})}\BibitemShut {NoStop}%
\bibitem [{\citenamefont {Barkett}\ \emph {et~al.}(2016)\citenamefont {Barkett}
  \emph {et~al.}}]{Barkett:2015wia}%
  \BibitemOpen
  \bibfield  {author} {\bibinfo {author} {\bibfnamefont {K.}~\bibnamefont
  {Barkett}} \emph {et~al.},\ }\href {\doibase 10.1103/PhysRevD.93.044064}
  {\bibfield  {journal} {\bibinfo  {journal} {Phys. Rev.}\ }\textbf {\bibinfo
  {volume} {D93}},\ \bibinfo {pages} {044064} (\bibinfo {year}
  {2016})}\BibitemShut {NoStop}%
\bibitem [{\citenamefont {Bernuzzi}\ \emph {et~al.}(2015)\citenamefont
  {Bernuzzi}, \citenamefont {Nagar}, \citenamefont {Dietrich},\ and\
  \citenamefont {Damour}}]{Bernuzzi:2014owa}%
  \BibitemOpen
  \bibfield  {author} {\bibinfo {author} {\bibfnamefont {S.}~\bibnamefont
  {Bernuzzi}}, \bibinfo {author} {\bibfnamefont {A.}~\bibnamefont {Nagar}},
  \bibinfo {author} {\bibfnamefont {T.}~\bibnamefont {Dietrich}}, \ and\
  \bibinfo {author} {\bibfnamefont {T.}~\bibnamefont {Damour}},\ }\href
  {\doibase 10.1103/PhysRevLett.114.161103} {\bibfield  {journal} {\bibinfo
  {journal} {Phys. Rev. Lett.}\ }\textbf {\bibinfo {volume} {114}},\ \bibinfo
  {pages} {161103} (\bibinfo {year} {2015})}\BibitemShut {NoStop}%
\bibitem [{\citenamefont {Hannam}\ \emph {et~al.}(2013)\citenamefont {Hannam},
  \citenamefont {Brown}, \citenamefont {Fairhurst}, \citenamefont {Fryer},\
  and\ \citenamefont {Harry}}]{Hannam:2013uu}%
  \BibitemOpen
  \bibfield  {author} {\bibinfo {author} {\bibfnamefont {M.}~\bibnamefont
  {Hannam}}, \bibinfo {author} {\bibfnamefont {D.~A.}\ \bibnamefont {Brown}},
  \bibinfo {author} {\bibfnamefont {S.}~\bibnamefont {Fairhurst}}, \bibinfo
  {author} {\bibfnamefont {C.~L.}\ \bibnamefont {Fryer}}, \ and\ \bibinfo
  {author} {\bibfnamefont {I.~W.}\ \bibnamefont {Harry}},\ }\href {\doibase
  10.1088/2041-8205/766/1/L14} {\bibfield  {journal} {\bibinfo  {journal}
  {Astrophys. J.}\ }\textbf {\bibinfo {volume} {766}},\ \bibinfo {pages} {L14}
  (\bibinfo {year} {2013})}\BibitemShut {NoStop}%
\bibitem [{\citenamefont {Lattimer}\ and\ \citenamefont
  {Lim}(2013)}]{Lattimer:2012xj}%
  \BibitemOpen
  \bibfield  {author} {\bibinfo {author} {\bibfnamefont {J.~M.}\ \bibnamefont
  {Lattimer}}\ and\ \bibinfo {author} {\bibfnamefont {Y.}~\bibnamefont {Lim}},\
  }\href {\doibase 10.1088/0004-637X/771/1/51} {\bibfield  {journal} {\bibinfo
  {journal} {Astrophys. J.}\ }\textbf {\bibinfo {volume} {771}},\ \bibinfo
  {pages} {51} (\bibinfo {year} {2013})}\BibitemShut {NoStop}%
\bibitem [{\citenamefont {Gandolfi}\ \emph {et~al.}(2012)\citenamefont
  {Gandolfi}, \citenamefont {Carlson},\ and\ \citenamefont
  {Reddy}}]{Gandolfi:2011xu}%
  \BibitemOpen
  \bibfield  {author} {\bibinfo {author} {\bibfnamefont {S.}~\bibnamefont
  {Gandolfi}}, \bibinfo {author} {\bibfnamefont {J.}~\bibnamefont {Carlson}}, \
  and\ \bibinfo {author} {\bibfnamefont {S.}~\bibnamefont {Reddy}},\ }\href
  {\doibase 10.1103/PhysRevC.85.032801} {\bibfield  {journal} {\bibinfo
  {journal} {Phys. Rev.}\ }\textbf {\bibinfo {volume} {C85}},\ \bibinfo {pages}
  {032801} (\bibinfo {year} {2012})}\BibitemShut {NoStop}%
\bibitem [{\citenamefont {Lynn}\ \emph {et~al.}(2016)\citenamefont {Lynn},
  \citenamefont {Tews}, \citenamefont {Carlson}, \citenamefont {Gandolfi},
  \citenamefont {Gezerlis}, \citenamefont {Schmidt},\ and\ \citenamefont
  {Schwenk}}]{Lynn:2015jua}%
  \BibitemOpen
  \bibfield  {author} {\bibinfo {author} {\bibfnamefont {J.~E.}\ \bibnamefont
  {Lynn}}, \bibinfo {author} {\bibfnamefont {I.}~\bibnamefont {Tews}}, \bibinfo
  {author} {\bibfnamefont {J.}~\bibnamefont {Carlson}}, \bibinfo {author}
  {\bibfnamefont {S.}~\bibnamefont {Gandolfi}}, \bibinfo {author}
  {\bibfnamefont {A.}~\bibnamefont {Gezerlis}}, \bibinfo {author}
  {\bibfnamefont {K.~E.}\ \bibnamefont {Schmidt}}, \ and\ \bibinfo {author}
  {\bibfnamefont {A.}~\bibnamefont {Schwenk}},\ }\href {\doibase
  10.1103/PhysRevLett.116.062501} {\bibfield  {journal} {\bibinfo  {journal}
  {Phys. Rev. Lett.}\ }\textbf {\bibinfo {volume} {116}},\ \bibinfo {pages}
  {062501} (\bibinfo {year} {2016})}\BibitemShut {NoStop}%
\bibitem [{\citenamefont {Drischler}\ \emph {et~al.}(2016)\citenamefont
  {Drischler}, \citenamefont {Hebeler},\ and\ \citenamefont
  {Schwenk}}]{Drischler:2015eba}%
  \BibitemOpen
  \bibfield  {author} {\bibinfo {author} {\bibfnamefont {C.}~\bibnamefont
  {Drischler}}, \bibinfo {author} {\bibfnamefont {K.}~\bibnamefont {Hebeler}},
  \ and\ \bibinfo {author} {\bibfnamefont {A.}~\bibnamefont {Schwenk}},\ }\href
  {\doibase 10.1103/PhysRevC.93.054314} {\bibfield  {journal} {\bibinfo
  {journal} {Phys. Rev.}\ }\textbf {\bibinfo {volume} {C93}},\ \bibinfo {pages}
  {054314} (\bibinfo {year} {2016})}\BibitemShut {NoStop}%
\bibitem [{\citenamefont {Tews}\ \emph {et~al.}(2017)\citenamefont {Tews},
  \citenamefont {Lattimer}, \citenamefont {Ohnishi},\ and\ \citenamefont
  {Kolomeitsev}}]{Kolomeitsev:2016sjl}%
  \BibitemOpen
  \bibfield  {author} {\bibinfo {author} {\bibfnamefont {I.}~\bibnamefont
  {Tews}}, \bibinfo {author} {\bibfnamefont {J.~M.}\ \bibnamefont {Lattimer}},
  \bibinfo {author} {\bibfnamefont {A.}~\bibnamefont {Ohnishi}}, \ and\
  \bibinfo {author} {\bibfnamefont {E.~E.}\ \bibnamefont {Kolomeitsev}},\
  }\href {\doibase 10.3847/1538-4357/aa8db9} {\bibfield  {journal} {\bibinfo
  {journal} {Astrophys. J.}\ }\textbf {\bibinfo {volume} {848}},\ \bibinfo
  {pages} {105} (\bibinfo {year} {2017})}\BibitemShut {NoStop}%
\bibitem [{\citenamefont {Tews}\ \emph {et~al.}(2013)\citenamefont {Tews},
  \citenamefont {Kruger}, \citenamefont {Hebeler},\ and\ \citenamefont
  {Schwenk}}]{Tews:2012fj}%
  \BibitemOpen
  \bibfield  {author} {\bibinfo {author} {\bibfnamefont {I.}~\bibnamefont
  {Tews}}, \bibinfo {author} {\bibfnamefont {T.}~\bibnamefont {Kr\"{u}ger}},
  \bibinfo {author} {\bibfnamefont {K.}~\bibnamefont {Hebeler}}, \ and\
  \bibinfo {author} {\bibfnamefont {A.}~\bibnamefont {Schwenk}},\ }\href
  {\doibase 10.1103/PhysRevLett.110.032504} {\bibfield  {journal} {\bibinfo
  {journal} {Phys. Rev. Lett.}\ }\textbf {\bibinfo {volume} {110}},\ \bibinfo
  {pages} {032504} (\bibinfo {year} {2013})}\BibitemShut {NoStop}%
\bibitem [{\citenamefont {Steiner}\ \emph {et~al.}(2010)\citenamefont
  {Steiner}, \citenamefont {Lattimer},\ and\ \citenamefont
  {Brown}}]{Steiner:2010fz}%
  \BibitemOpen
  \bibfield  {author} {\bibinfo {author} {\bibfnamefont {A.~W.}\ \bibnamefont
  {Steiner}}, \bibinfo {author} {\bibfnamefont {J.~M.}\ \bibnamefont
  {Lattimer}}, \ and\ \bibinfo {author} {\bibfnamefont {E.~F.}\ \bibnamefont
  {Brown}},\ }\href {\doibase 10.1088/0004-637X/722/1/33} {\bibfield  {journal}
  {\bibinfo  {journal} {Astrophys. J.}\ }\textbf {\bibinfo {volume} {722}},\
  \bibinfo {pages} {33} (\bibinfo {year} {2010})}\BibitemShut {NoStop}%
\bibitem [{\citenamefont {Degenaar}\ and\ \citenamefont
  {Suleimanov}(2018)}]{Degenaar:2018lle}%
  \BibitemOpen
  \bibfield  {author} {\bibinfo {author} {\bibfnamefont {N.}~\bibnamefont
  {Degenaar}}\ and\ \bibinfo {author} {\bibfnamefont {V.~F.}\ \bibnamefont
  {Suleimanov}},\ }\href@noop {} {\  (\bibinfo {year} {2018})},\ \Eprint
  {http://arxiv.org/abs/1806.02833} {arXiv:1806.02833 [astro-ph.HE]}
  \BibitemShut {NoStop}%
\bibitem [{\citenamefont {Guillot}\ and\ \citenamefont
  {Rutledge}(2014)}]{Guillot:2014lla}%
  \BibitemOpen
  \bibfield  {author} {\bibinfo {author} {\bibfnamefont {S.}~\bibnamefont
  {Guillot}}\ and\ \bibinfo {author} {\bibfnamefont {R.~E.}\ \bibnamefont
  {Rutledge}},\ }\href {\doibase 10.1088/2041-8205/796/1/L3} {\bibfield
  {journal} {\bibinfo  {journal} {Astrophys. J.}\ }\textbf {\bibinfo {volume}
  {796}},\ \bibinfo {pages} {L3} (\bibinfo {year} {2014})}\BibitemShut
  {NoStop}%
\bibitem [{\citenamefont {Lattimer}\ and\ \citenamefont
  {Steiner}(2014)}]{Lattimer:2013hma}%
  \BibitemOpen
  \bibfield  {author} {\bibinfo {author} {\bibfnamefont {J.~M.}\ \bibnamefont
  {Lattimer}}\ and\ \bibinfo {author} {\bibfnamefont {A.~W.}\ \bibnamefont
  {Steiner}},\ }\href {\doibase 10.1088/0004-637X/784/2/123} {\bibfield
  {journal} {\bibinfo  {journal} {Astrophys. J.}\ }\textbf {\bibinfo {volume}
  {784}},\ \bibinfo {pages} {123} (\bibinfo {year} {2014})}\BibitemShut
  {NoStop}%
\bibitem [{\citenamefont {Shaw}\ \emph {et~al.}(2018)\citenamefont {Shaw},
  \citenamefont {Heinke}, \citenamefont {Steiner}, \citenamefont {Campana},
  \citenamefont {Cohn}, \citenamefont {Ho}, \citenamefont {Lugger},\ and\
  \citenamefont {Servillat}}]{Shaw:2018wxh}%
  \BibitemOpen
  \bibfield  {author} {\bibinfo {author} {\bibfnamefont {A.~W.}\ \bibnamefont
  {Shaw}}, \bibinfo {author} {\bibfnamefont {C.~O.}\ \bibnamefont {Heinke}},
  \bibinfo {author} {\bibfnamefont {A.~W.}\ \bibnamefont {Steiner}}, \bibinfo
  {author} {\bibfnamefont {S.}~\bibnamefont {Campana}}, \bibinfo {author}
  {\bibfnamefont {H.~N.}\ \bibnamefont {Cohn}}, \bibinfo {author}
  {\bibfnamefont {W.~C.~G.}\ \bibnamefont {Ho}}, \bibinfo {author}
  {\bibfnamefont {P.~M.}\ \bibnamefont {Lugger}}, \ and\ \bibinfo {author}
  {\bibfnamefont {M.}~\bibnamefont {Servillat}},\ }\href {\doibase 10.1093/mnras/sty582} {\bibfield
  {journal} {\bibinfo  {journal} {Mon. Not. Roy. Astron. Soc.}\ }\textbf {\bibinfo {volume}
  {476}},\ \bibinfo {pages} {4713} (\bibinfo {year} {2018})} \BibitemShut {NoStop}%
\bibitem [{\citenamefont {De}\ \emph {et~al.}(2018)\citenamefont {De}
  \emph {et~al.}}]{De:2018prep}%
  \BibitemOpen
  \bibfield  {author} {\bibinfo {author} {\bibfnamefont {S.}~\bibnamefont
  {De}} \emph {et~al.},\ }
  {\bibfield  {journal} {\bibinfo  {journal} {(in preparation)}\ }
  }\BibitemShut {NoStop}%
\bibitem [{\citenamefont {Abbott}\ \emph {et~al.}(2018)\citenamefont {Abbott}
  \emph {et~al.}}]{Abbott:2018exr}%
  \BibitemOpen
  \bibfield  {author} {\bibinfo {author} {\bibfnamefont {B.~P.}\ \bibnamefont
  {Abbott}} \emph {et~al.} (\bibinfo {collaboration} {Virgo, LIGO
  Scientific}),\ }\href@noop {} {\  (\bibinfo {year} {2018})},\ \Eprint
  {http://arxiv.org/abs/1805.11581} {arXiv:1805.11581 [gr-qc]} \BibitemShut
  {NoStop}%
\bibitem [{\citenamefont {Yagi}\ and\ \citenamefont
  {Yunes}(2017)}]{Yagi:2016qmr}%
  \BibitemOpen
  \bibfield  {author} {\bibinfo {author} {\bibfnamefont {K.}~\bibnamefont
  {Yagi}}\ and\ \bibinfo {author} {\bibfnamefont {N.}~\bibnamefont {Yunes}},\
  }\href {\doibase 10.1088/1361-6382/34/1/015006} {\bibfield  {journal}
  {\bibinfo  {journal} {Class. Quant. Grav.}\ }\textbf {\bibinfo {volume}
  {34}},\ \bibinfo {pages} {015006} (\bibinfo {year} {2017})}\BibitemShut
  {NoStop}%
\bibitem [{\citenamefont {Chatziioannou}\ \emph {et~al.}(2018)\citenamefont
  {Chatziioannou}, \citenamefont {Haster},\ and\ \citenamefont
  {Zimmerman}}]{Chatziioannou:2018vzf}%
  \BibitemOpen
  \bibfield  {author} {\bibinfo {author} {\bibfnamefont {K.}~\bibnamefont
  {Chatziioannou}}, \bibinfo {author} {\bibfnamefont {C.-J.}\ \bibnamefont
  {Haster}}, \ and\ \bibinfo {author} {\bibfnamefont {A.}~\bibnamefont
  {Zimmerman}},\ }\href {\doibase 10.1103/PhysRevD.97.104036} {\bibfield
  {journal} {\bibinfo  {journal} {Phys. Rev.}\ }\textbf {\bibinfo {volume}
  {D97}},\ \bibinfo {pages} {104036} (\bibinfo {year} {2018})}\BibitemShut
  {NoStop}%


\end{thebibliography}

\begin{thebibliography}{31}%
\makeatletter
\providecommand \@ifxundefined [1]{%
 \@ifx{#1\undefined}
}%
\providecommand \@ifnum [1]{%
 \ifnum #1\expandafter \@firstoftwo
 \else \expandafter \@secondoftwo
 \fi
}%
\providecommand \@ifx [1]{%
 \ifx #1\expandafter \@firstoftwo
 \else \expandafter \@secondoftwo
 \fi
}%
\providecommand \natexlab [1]{#1}%
\providecommand \enquote  [1]{``#1''}%
\providecommand \bibnamefont  [1]{#1}%
\providecommand \bibfnamefont [1]{#1}%
\providecommand \citenamefont [1]{#1}%
\providecommand \href@noop [0]{\@secondoftwo}%
\providecommand \href [0]{\begingroup \@sanitize@url \@href}%
\providecommand \@href[1]{\@@startlink{#1}\@@href}%
\providecommand \@@href[1]{\endgroup#1\@@endlink}%
\providecommand \@sanitize@url [0]{\catcode `\\12\catcode `\$12\catcode
  `\&12\catcode `\#12\catcode `\^12\catcode `\_12\catcode `\%12\relax}%
\providecommand \@@startlink[1]{}%
\providecommand \@@endlink[0]{}%
\providecommand \url  [0]{\begingroup\@sanitize@url \@url }%
\providecommand \@url [1]{\endgroup\@href {#1}{\urlprefix }}%
\providecommand \urlprefix  [0]{URL }%
\providecommand \Eprint [0]{\href }%
\providecommand \doibase [0]{http://dx.doi.org/}%
\providecommand \selectlanguage [0]{\@gobble}%
\providecommand \bibinfo  [0]{\@secondoftwo}%
\providecommand \bibfield  [0]{\@secondoftwo}%
\providecommand \translation [1]{[#1]}%
\providecommand \BibitemOpen [0]{}%
\providecommand \bibitemStop [0]{}%
\providecommand \bibitemNoStop [0]{.\EOS\space}%
\providecommand \EOS [0]{\spacefactor3000\relax}%
\providecommand \BibitemShut  [1]{\csname bibitem#1\endcsname}%
\let\auto@bib@innerbib\@empty
\bibitem [{\citenamefont {Oppenheimer}\ and\ \citenamefont
  {Volkoff}(1939)}]{Oppenheimer:1939ne}%
  \BibitemOpen
  \bibfield  {author} {\bibinfo {author} {\bibfnamefont {J.~R.}\ \bibnamefont
  {Oppenheimer}}\ and\ \bibinfo {author} {\bibfnamefont {G.~M.}\ \bibnamefont
  {Volkoff}},\ }\href {\doibase 10.1103/PhysRev.55.374} {\bibfield  {journal}
  {\bibinfo  {journal} {Phys. Rev.}\ }\textbf {\bibinfo {volume} {55}},\
  \bibinfo {pages} {374} (\bibinfo {year} {1939})}\BibitemShut {NoStop}%
\bibitem [{\citenamefont {Lattimer}\ and\ \citenamefont
  {Prakash}(2016)}]{Lattimer:2015nhk}%
  \BibitemOpen
  \bibfield  {author} {\bibinfo {author} {\bibfnamefont {J.~M.}\ \bibnamefont
  {Lattimer}}\ and\ \bibinfo {author} {\bibfnamefont {M.}~\bibnamefont
  {Prakash}},\ }\href {\doibase 10.1016/j.physrep.2015.12.005} {\bibfield
  {journal} {\bibinfo  {journal} {Phys. Rept.}\ }\textbf {\bibinfo {volume}
  {621}},\ \bibinfo {pages} {127} (\bibinfo {year} {2016})}\BibitemShut
  {NoStop}%
\bibitem [{\citenamefont {Tauris}\ \emph {et~al.}(2017)\citenamefont {Tauris}
  \emph {et~al.}}]{Tauris:2017omb}%
  \BibitemOpen
  \bibfield  {author} {\bibinfo {author} {\bibfnamefont {T.~M.}\ \bibnamefont
  {Tauris}} \emph {et~al.},\ }\href {\doibase 10.3847/1538-4357/aa7e89}
  {\bibfield  {journal} {\bibinfo  {journal} {Astrophys. J.}\ }\textbf
  {\bibinfo {volume} {846}},\ \bibinfo {pages} {170} (\bibinfo {year}
  {2017})}\BibitemShut {NoStop}%
\bibitem [{\citenamefont {\"{O}zel}\ and\ \citenamefont
  {Freire}(2016)}]{Ozel:2016oaf}%
  \BibitemOpen
  \bibfield  {author} {\bibinfo {author} {\bibfnamefont {F.}~\bibnamefont
  {\"{O}zel}}\ and\ \bibinfo {author} {\bibfnamefont {P.}~\bibnamefont
  {Freire}},\ }\href {\doibase 10.1146/annurev-astro-081915-023322} {\bibfield
  {journal} {\bibinfo  {journal} {Ann. Rev. Astron. Astrophys.}\ }\textbf
  {\bibinfo {volume} {54}},\ \bibinfo {pages} {401} (\bibinfo {year}
  {2016})}\BibitemShut {NoStop}%
\bibitem [{\citenamefont {Lattimer}(2012)}]{Lattimer:2012nd}%
  \BibitemOpen
  \bibfield  {author} {\bibinfo {author} {\bibfnamefont {J.~M.}\ \bibnamefont
  {Lattimer}},\ }\href {\doibase 10.1146/annurev-nucl-102711-095018} {\bibfield
   {journal} {\bibinfo  {journal} {Ann. Rev. Nucl. Part. Sci.}\ }\textbf
  {\bibinfo {volume} {62}},\ \bibinfo {pages} {485} (\bibinfo {year}
  {2012})}\BibitemShut {NoStop}%
\bibitem [{\citenamefont {Zhao}\ and\ \citenamefont
  {Lattimer}(2018)}]{Zhao:2018nyf}%
  \BibitemOpen
  \bibfield  {author} {\bibinfo {author} {\bibfnamefont {T.}~\bibnamefont
  {Zhao}}\ and\ \bibinfo {author} {\bibfnamefont {J.~M.}\ \bibnamefont
  {Lattimer}},\ }\href@noop {} {\  (\bibinfo {year} {2018})},\ \Eprint
  {http://arxiv.org/abs/1808.02858} {arXiv:1808.02858 [astro-ph.HE]}
  \BibitemShut {NoStop}%
\bibitem [{\citenamefont {Vallisneri}\ \emph {et~al.}(2015)\citenamefont
  {Vallisneri}, \citenamefont {Kanner}, \citenamefont {Williams}, \citenamefont
  {Weinstein},\ and\ \citenamefont {Stephens}}]{Vallisneri:2014vxa}%
  \BibitemOpen
  \bibfield  {author} {\bibinfo {author} {\bibfnamefont {M.}~\bibnamefont
  {Vallisneri}}, \bibinfo {author} {\bibfnamefont {J.}~\bibnamefont {Kanner}},
  \bibinfo {author} {\bibfnamefont {R.}~\bibnamefont {Williams}}, \bibinfo
  {author} {\bibfnamefont {A.}~\bibnamefont {Weinstein}}, \ and\ \bibinfo
  {author} {\bibfnamefont {B.}~\bibnamefont {Stephens}},\ }\href
  {http://stacks.iop.org/1742-6596/610/i=1/a=012021} {\bibfield  {journal}
  {\bibinfo  {journal} {Journal of Physics: Conference Series}\ }\textbf
  {\bibinfo {volume} {610}},\ \bibinfo {pages} {012021} (\bibinfo {year}
  {2015})}\BibitemShut {NoStop}%
\bibitem [{\citenamefont {Blackburn}\ \emph {et~al.}(2017)\citenamefont
  {Blackburn} \emph {et~al.}}]{gw170817-losc}%
  \BibitemOpen
  \bibfield  {author} {\bibinfo {author} {\bibfnamefont {K.}~\bibnamefont
  {Blackburn}} \emph {et~al.},\ }\href {\doibase doi:10.7935/K5B8566F} {\emph
  {\bibinfo {title} {{LOSC CLN Data Products for GW170817}}}} (\bibinfo {year}
  {2017})\BibitemShut {NoStop}%
\bibitem [{\citenamefont {Biwer}\ \emph {et~al.}(2018)\citenamefont {Biwer},
  \citenamefont {Capano}, \citenamefont {De}, \citenamefont {Cabero},
  \citenamefont {Brown}, \citenamefont {Nitz},\ and\ \citenamefont
  {Raymond}}]{Biwer:2018osg}%
  \BibitemOpen
  \bibfield  {author} {\bibinfo {author} {\bibfnamefont {C.~M.}\ \bibnamefont
  {Biwer}}, \bibinfo {author} {\bibfnamefont {C.~D.}\ \bibnamefont {Capano}},
  \bibinfo {author} {\bibfnamefont {S.}~\bibnamefont {De}}, \bibinfo {author}
  {\bibfnamefont {M.}~\bibnamefont {Cabero}}, \bibinfo {author} {\bibfnamefont
  {D.~A.}\ \bibnamefont {Brown}}, \bibinfo {author} {\bibfnamefont {A.~H.}\
  \bibnamefont {Nitz}}, \ and\ \bibinfo {author} {\bibfnamefont
  {V.}~\bibnamefont {Raymond}},\ }\href@noop {} {\  (\bibinfo {year} {2018})},\
  \Eprint {http://arxiv.org/abs/1807.10312} {arXiv:1807.10312 [astro-ph.IM]}
  \BibitemShut {NoStop}%
\bibitem [{\citenamefont {Nitz}\ \emph {et~al.}(2018)\citenamefont {Nitz} \emph
  {et~al.}}]{alex_nitz_2018_1208115}%
  \BibitemOpen
  \bibfield  {author} {\bibinfo {author} {\bibfnamefont {A.}~\bibnamefont
  {Nitz}} \emph {et~al.},\ }\href {\doibase 10.5281/zenodo.1208115} {\emph
  {\bibinfo {title} {PyCBC v1.9.4}}} (\bibinfo {year} {2018})\BibitemShut
  {NoStop}%
\bibitem [{\citenamefont {Foreman-Mackey}\ \emph {et~al.}(2013)\citenamefont
  {Foreman-Mackey}, \citenamefont {Hogg}, \citenamefont {Lang},\ and\
  \citenamefont {Goodman}}]{emcee}%
  \BibitemOpen
  \bibfield  {author} {\bibinfo {author} {\bibfnamefont {D.}~\bibnamefont
  {Foreman-Mackey}}, \bibinfo {author} {\bibfnamefont {D.~W.}\ \bibnamefont
  {Hogg}}, \bibinfo {author} {\bibfnamefont {D.}~\bibnamefont {Lang}}, \ and\
  \bibinfo {author} {\bibfnamefont {J.}~\bibnamefont {Goodman}},\ }\href
  {http://stacks.iop.org/1538-3873/125/i=925/a=306} {\bibfield  {journal}
  {\bibinfo  {journal} {Publications of the Astronomical Society of the
  Pacific}\ }\textbf {\bibinfo {volume} {125}},\ \bibinfo {pages} {306}
  (\bibinfo {year} {2013})}\BibitemShut {NoStop}%
\bibitem [{\citenamefont {Vousden}\ \emph {et~al.}(2016)\citenamefont
  {Vousden}, \citenamefont {Farr},\ and\ \citenamefont
  {Mandel}}]{vousden:2016}%
  \BibitemOpen
  \bibfield  {author} {\bibinfo {author} {\bibfnamefont {W.~D.}\ \bibnamefont
  {Vousden}}, \bibinfo {author} {\bibfnamefont {W.~M.}\ \bibnamefont {Farr}}, \
  and\ \bibinfo {author} {\bibfnamefont {I.}~\bibnamefont {Mandel}},\ }\href
  {\doibase 10.1093/mnras/stv2422} {\bibfield  {journal} {\bibinfo  {journal}
  {Monthly Notices of the Royal Astronomical Society}\ }\textbf {\bibinfo
  {volume} {455}},\ \bibinfo {pages} {1919} (\bibinfo {year}
  {2016})}\BibitemShut {NoStop}%
\bibitem [{\citenamefont {Goodman~J.}(2010)}]{mcmc}%
  \BibitemOpen
  \bibfield  {author} {\bibinfo {author} {\bibfnamefont {W.~J.}\ \bibnamefont
  {Goodman~J.}},\ }\href {\doibase 10.2140/camcos.2010.5.65} {\bibfield
  {journal} {\bibinfo  {journal} {Commun. Appl. Math. Comput. Sci.}\ }\textbf
  {\bibinfo {volume} {5}},\ \bibinfo {pages} {65} (\bibinfo {year}
  {2010})}\BibitemShut {NoStop}%
\bibitem [{\citenamefont {Driggers}\ \emph {et~al.}(2017)\citenamefont
  {Driggers}, \citenamefont {Vitale}, \citenamefont {Lundgren}, \citenamefont
  {Evans}, \citenamefont {Kawabe}, \citenamefont {Dwyer}, \citenamefont
  {Izumi},\ and\ \citenamefont {Fritschel}}]{gw170817-noise}%
  \BibitemOpen
  \bibfield  {author} {\bibinfo {author} {\bibfnamefont {J.}~\bibnamefont
  {Driggers}}, \bibinfo {author} {\bibfnamefont {S.}~\bibnamefont {Vitale}},
  \bibinfo {author} {\bibfnamefont {A.}~\bibnamefont {Lundgren}}, \bibinfo
  {author} {\bibfnamefont {M.}~\bibnamefont {Evans}}, \bibinfo {author}
  {\bibfnamefont {K.}~\bibnamefont {Kawabe}}, \bibinfo {author} {\bibfnamefont
  {S.}~\bibnamefont {Dwyer}}, \bibinfo {author} {\bibfnamefont
  {K.}~\bibnamefont {Izumi}}, \ and\ \bibinfo {author} {\bibfnamefont
  {P.}~\bibnamefont {Fritschel}},\ }\href@noop {} {\enquote {\bibinfo {title}
  {Offline noise subtraction for {Advanced LIGO}},}\ } (\bibinfo {year}
  {2017}),\ \bibinfo {note}
  {{https://dcc.ligo.org/LIGO-P1700260/public}}\BibitemShut {NoStop}%
\bibitem [{\citenamefont {Welch}(1967)}]{1161901}%
  \BibitemOpen
  \bibfield  {author} {\bibinfo {author} {\bibfnamefont {P.}~\bibnamefont
  {Welch}},\ }\href {\doibase 10.1109/TAU.1967.1161901} {\bibfield  {journal}
  {\bibinfo  {journal} {IEEE Transactions on Audio and Electroacoustics}\
  }\textbf {\bibinfo {volume} {15}},\ \bibinfo {pages} {70} (\bibinfo {year}
  {1967})}\BibitemShut {NoStop}%
\bibitem [{\citenamefont {Allen}\ \emph {et~al.}(2012)\citenamefont {Allen},
  \citenamefont {Anderson}, \citenamefont {Brady}, \citenamefont {Brown},\ and\
  \citenamefont {Creighton}}]{Allen:2005fk}%
  \BibitemOpen
  \bibfield  {author} {\bibinfo {author} {\bibfnamefont {B.}~\bibnamefont
  {Allen}}, \bibinfo {author} {\bibfnamefont {W.~G.}\ \bibnamefont {Anderson}},
  \bibinfo {author} {\bibfnamefont {P.~R.}\ \bibnamefont {Brady}}, \bibinfo
  {author} {\bibfnamefont {D.~A.}\ \bibnamefont {Brown}}, \ and\ \bibinfo
  {author} {\bibfnamefont {J.~D.~E.}\ \bibnamefont {Creighton}},\ }\href
  {\doibase 10.1103/PhysRevD.85.122006} {\bibfield  {journal} {\bibinfo
  {journal} {Phys. Rev.}\ }\textbf {\bibinfo {volume} {D85}},\ \bibinfo {pages}
  {122006} (\bibinfo {year} {2012})}\BibitemShut {NoStop}%
\bibitem [{\citenamefont {Abbott}\ \emph {et~al.}(2018)\citenamefont {Abbott}
  \emph {et~al.}}]{Abbott:2018wiz}%
  \BibitemOpen
  \bibfield  {author} {\bibinfo {author} {\bibfnamefont {B.~P.}\ \bibnamefont
  {Abbott}} \emph {et~al.} (\bibinfo {collaboration} {Virgo, LIGO
  Scientific}),\ }\href@noop {} {\  (\bibinfo {year} {2018})},\ \Eprint
  {http://arxiv.org/abs/1805.11579} {arXiv:1805.11579 [gr-qc]} \BibitemShut
  {NoStop}%
\bibitem [{\citenamefont {Littenberg}\ and\ \citenamefont
  {Cornish}(2015)}]{Littenberg:2014oda}%
  \BibitemOpen
  \bibfield  {author} {\bibinfo {author} {\bibfnamefont {T.~B.}\ \bibnamefont
  {Littenberg}}\ and\ \bibinfo {author} {\bibfnamefont {N.~J.}\ \bibnamefont
  {Cornish}},\ }\href {\doibase 10.1103/PhysRevD.91.084034} {\bibfield
  {journal} {\bibinfo  {journal} {Phys. Rev.}\ }\textbf {\bibinfo {volume}
  {D91}},\ \bibinfo {pages} {084034} (\bibinfo {year} {2015})}\BibitemShut
  {NoStop}%
\bibitem [{\citenamefont {Damour}\ \emph {et~al.}(2012)\citenamefont {Damour},
  \citenamefont {Nagar},\ and\ \citenamefont {Villain}}]{Damour:2012yf}%
  \BibitemOpen
  \bibfield  {author} {\bibinfo {author} {\bibfnamefont {T.}~\bibnamefont
  {Damour}}, \bibinfo {author} {\bibfnamefont {A.}~\bibnamefont {Nagar}}, \
  and\ \bibinfo {author} {\bibfnamefont {L.}~\bibnamefont {Villain}},\ }\href
  {\doibase 10.1103/PhysRevD.85.123007} {\bibfield  {journal} {\bibinfo
  {journal} {Phys. Rev.}\ }\textbf {\bibinfo {volume} {D85}},\ \bibinfo {pages}
  {123007} (\bibinfo {year} {2012})}\BibitemShut {NoStop}%
\bibitem [{\citenamefont {Mercer}\ \emph {et~al.}(2017)\citenamefont {Mercer}
  \emph {et~al.}}]{lal}%
  \BibitemOpen
  \bibfield  {author} {\bibinfo {author} {\bibfnamefont {R.~A.}\ \bibnamefont
  {Mercer}} \emph {et~al.},\ }\href@noop {} {\enquote {\bibinfo {title} {{LIGO
  Algorithm Library}},}\ } (\bibinfo {year} {2017}),\ \bibinfo {note}
  {{https://git.ligo.org/lscsoft/lalsuite}}\BibitemShut {NoStop}%
\bibitem [{\citenamefont {Buonanno}\ \emph {et~al.}(2009)\citenamefont
  {Buonanno}, \citenamefont {Iyer}, \citenamefont {Ochsner}, \citenamefont
  {Pan},\ and\ \citenamefont {Sathyaprakash}}]{Buonanno:2009zt}%
  \BibitemOpen
  \bibfield  {author} {\bibinfo {author} {\bibfnamefont {A.}~\bibnamefont
  {Buonanno}}, \bibinfo {author} {\bibfnamefont {B.~R.}\ \bibnamefont {Iyer}},
  \bibinfo {author} {\bibfnamefont {E.}~\bibnamefont {Ochsner}}, \bibinfo
  {author} {\bibfnamefont {Y.}~\bibnamefont {Pan}}, \ and\ \bibinfo {author}
  {\bibfnamefont {B.~S.}\ \bibnamefont {Sathyaprakash}},\ }\href {\doibase
  10.1103/PhysRevD.80.084043} {\bibfield  {journal} {\bibinfo  {journal} {Phys.
  Rev. D}\ }\textbf {\bibinfo {volume} {80}},\ \bibinfo {pages} {084043}
  (\bibinfo {year} {2009})}\BibitemShut {NoStop}%
\bibitem [{\citenamefont {Arun}\ \emph {et~al.}(2009)\citenamefont {Arun},
  \citenamefont {Buonanno}, \citenamefont {Faye},\ and\ \citenamefont
  {Ochsner}}]{Arun:2008kb}%
  \BibitemOpen
  \bibfield  {author} {\bibinfo {author} {\bibfnamefont {K.~G.}\ \bibnamefont
  {Arun}}, \bibinfo {author} {\bibfnamefont {A.}~\bibnamefont {Buonanno}},
  \bibinfo {author} {\bibfnamefont {G.}~\bibnamefont {Faye}}, \ and\ \bibinfo
  {author} {\bibfnamefont {E.}~\bibnamefont {Ochsner}},\ }\href {\doibase
  10.1103/PhysRevD.79.104023, 10.1103/PhysRevD.84.049901} {\bibfield  {journal}
  {\bibinfo  {journal} {Phys. Rev.}\ }\textbf {\bibinfo {volume} {D79}},\
  \bibinfo {pages} {104023} (\bibinfo {year} {2009})},\ \bibinfo {note}
  {[Erratum: Phys. Rev.D84,049901(2011)]}\BibitemShut {NoStop}%
\bibitem [{\citenamefont {Mikoczi}\ \emph {et~al.}(2005)\citenamefont
  {Mikoczi}, \citenamefont {Vasuth},\ and\ \citenamefont
  {Gergely}}]{Mikoczi:2005dn}%
  \BibitemOpen
  \bibfield  {author} {\bibinfo {author} {\bibfnamefont {B.}~\bibnamefont
  {Mikoczi}}, \bibinfo {author} {\bibfnamefont {M.}~\bibnamefont {Vasuth}}, \
  and\ \bibinfo {author} {\bibfnamefont {L.~A.}\ \bibnamefont {Gergely}},\
  }\href {\doibase 10.1103/PhysRevD.71.124043} {\bibfield  {journal} {\bibinfo
  {journal} {Phys. Rev.}\ }\textbf {\bibinfo {volume} {D71}},\ \bibinfo {pages}
  {124043} (\bibinfo {year} {2005})}\BibitemShut {NoStop}%
\bibitem [{\citenamefont {Boh\'e}\ \emph {et~al.}(2013)\citenamefont {Boh\'e},
  \citenamefont {Marsat},\ and\ \citenamefont {Blanchet}}]{Bohe:2013cla}%
  \BibitemOpen
  \bibfield  {author} {\bibinfo {author} {\bibfnamefont {A.}~\bibnamefont
  {Boh\'e}}, \bibinfo {author} {\bibfnamefont {S.}~\bibnamefont {Marsat}}, \
  and\ \bibinfo {author} {\bibfnamefont {L.}~\bibnamefont {Blanchet}},\ }\href
  {http://stacks.iop.org/0264-9381/30/i=13/a=135009} {\bibfield  {journal}
  {\bibinfo  {journal} {Classical and Quantum Gravity}\ }\textbf {\bibinfo
  {volume} {30}},\ \bibinfo {pages} {135009} (\bibinfo {year}
  {2013})}\BibitemShut {NoStop}%
\bibitem [{\citenamefont {Vines}\ \emph {et~al.}(2011)\citenamefont {Vines},
  \citenamefont {Flanagan},\ and\ \citenamefont {Hinderer}}]{Vines:2011ud}%
  \BibitemOpen
  \bibfield  {author} {\bibinfo {author} {\bibfnamefont {J.}~\bibnamefont
  {Vines}}, \bibinfo {author} {\bibfnamefont {\'{E}.~\'{E}.}\ \bibnamefont {Flanagan}},
  \ and\ \bibinfo {author} {\bibfnamefont {T.}~\bibnamefont {Hinderer}},\
  }\href {\doibase 10.1103/PhysRevD.83.084051} {\bibfield  {journal} {\bibinfo
  {journal} {Phys. Rev. D}\ }\textbf {\bibinfo {volume} {83}},\ \bibinfo
  {pages} {084051} (\bibinfo {year} {2011})}\BibitemShut {NoStop}%
\bibitem [{\citenamefont {Brown}\ \emph {et~al.}(2012)\citenamefont {Brown},
  \citenamefont {Harry}, \citenamefont {Lundgren},\ and\ \citenamefont
  {Nitz}}]{Brown:2012qf}%
  \BibitemOpen
  \bibfield  {author} {\bibinfo {author} {\bibfnamefont {D.~A.}\ \bibnamefont
  {Brown}}, \bibinfo {author} {\bibfnamefont {I.}~\bibnamefont {Harry}},
  \bibinfo {author} {\bibfnamefont {A.}~\bibnamefont {Lundgren}}, \ and\
  \bibinfo {author} {\bibfnamefont {A.~H.}\ \bibnamefont {Nitz}},\ }\href
  {\doibase 10.1103/PhysRevD.86.084017} {\bibfield  {journal} {\bibinfo
  {journal} {Phys. Rev. D}\ }\textbf {\bibinfo {volume} {86}},\ \bibinfo
  {pages} {084017} (\bibinfo {year} {2012})}\BibitemShut {NoStop}%
\bibitem [{\citenamefont {Soares-Santos}\ \emph {et~al.}(2017)\citenamefont
  {Soares-Santos} \emph {et~al.}}]{Soares-Santos:2017lru}%
  \BibitemOpen
  \bibfield  {author} {\bibinfo {author} {\bibfnamefont {M.}~\bibnamefont
  {Soares-Santos}} \emph {et~al.},\ }\href {\doibase 10.3847/2041-8213/aa9059}
  {\bibfield  {journal} {\bibinfo  {journal} {Astrophys. J.}\ }\textbf
  {\bibinfo {volume} {848}},\ \bibinfo {pages} {L16} (\bibinfo {year}
  {2017})}\BibitemShut {NoStop}%
\bibitem [{\citenamefont {Cantiello}\ \emph {et~al.}(2018)\citenamefont
  {Cantiello} \emph {et~al.}}]{Cantiello:2018ffy}%
  \BibitemOpen
  \bibfield  {author} {\bibinfo {author} {\bibfnamefont {M.}~\bibnamefont
  {Cantiello}} \emph {et~al.},\ }\href {\doibase 10.3847/2041-8213/aaad64}
  {\bibfield  {journal} {\bibinfo  {journal} {Astrophys. J.}\ }\textbf
  {\bibinfo {volume} {854}},\ \bibinfo {pages} {L31} (\bibinfo {year}
  {2018})}\BibitemShut {NoStop}%
\bibitem [{\citenamefont {Abbott}\ \emph {et~al.}(2017)\citenamefont {Abbott}
  \emph {et~al.}}]{TheLIGOScientific:2017qsa}%
  \BibitemOpen
  \bibfield  {author} {\bibinfo {author} {\bibfnamefont {B.}~\bibnamefont
  {Abbott}} \emph {et~al.},\ }\href {\doibase 10.1103/PhysRevLett.119.161101}
  {\bibfield  {journal} {\bibinfo  {journal} {Phys. Rev. Lett.}\ }\textbf
  {\bibinfo {volume} {119}},\ \bibinfo {pages} {161101} (\bibinfo {year}
  {2017})}\BibitemShut {NoStop}%
\bibitem [{\citenamefont {P\'erez}\ and\ \citenamefont
  {Granger}(2007)}]{PER-GRA:2007}%
  \BibitemOpen
  \bibfield  {author} {\bibinfo {author} {\bibfnamefont {F.}~\bibnamefont
  {P\'erez}}\ and\ \bibinfo {author} {\bibfnamefont {B.~E.}\ \bibnamefont
  {Granger}},\ }\href {\doibase 10.1109/MCSE.2007.53} {\bibfield  {journal}
  {\bibinfo  {journal} {Computing in Science and Engineering}\ }\textbf
  {\bibinfo {volume} {9}},\ \bibinfo {pages} {21} (\bibinfo {year}
  {2007})}\BibitemShut {NoStop}%
\bibitem [{\citenamefont {De}\ \emph {et~al.}(2018)\citenamefont {De} \emph
  {et~al.}}]{gw170817commoneos}%
  \BibitemOpen
  \bibfield  {author} {\bibinfo {author} {\bibfnamefont {S.}~\bibnamefont {De}}
  \emph {et~al.},\ }\href {https://github.com/sugwg/gw170817-common-eos} {\emph
  {\bibinfo {title} {SUGWG GitHub Repository}}} (\bibinfo {year}
  {2018})\BibitemShut {NoStop}%
\end{thebibliography}

\begin{thebibliography}{6}%
\makeatletter
\providecommand \@ifxundefined [1]{%
 \@ifx{#1\undefined}
}%
\providecommand \@ifnum [1]{%
 \ifnum #1\expandafter \@firstoftwo
 \else \expandafter \@secondoftwo
 \fi
}%
\providecommand \@ifx [1]{%
 \ifx #1\expandafter \@firstoftwo
 \else \expandafter \@secondoftwo
 \fi
}%
\providecommand \natexlab [1]{#1}%
\providecommand \enquote  [1]{``#1''}%
\providecommand \bibnamefont  [1]{#1}%
\providecommand \bibfnamefont [1]{#1}%
\providecommand \citenamefont [1]{#1}%
\providecommand \href@noop [0]{\@secondoftwo}%
\providecommand \href [0]{\begingroup \@sanitize@url \@href}%
\providecommand \@href[1]{\@@startlink{#1}\@@href}%
\providecommand \@@href[1]{\endgroup#1\@@endlink}%
\providecommand \@sanitize@url [0]{\catcode `\\12\catcode `\$12\catcode
  `\&12\catcode `\#12\catcode `\^12\catcode `\_12\catcode `\%12\relax}%
\providecommand \@@startlink[1]{}%
\providecommand \@@endlink[0]{}%
\providecommand \url  [0]{\begingroup\@sanitize@url \@url }%
\providecommand \@url [1]{\endgroup\@href {#1}{\urlprefix }}%
\providecommand \urlprefix  [0]{URL }%
\providecommand \Eprint [0]{\href }%
\providecommand \doibase [0]{http://dx.doi.org/}%
\providecommand \selectlanguage [0]{\@gobble}%
\providecommand \bibinfo  [0]{\@secondoftwo}%
\providecommand \bibfield  [0]{\@secondoftwo}%
\providecommand \translation [1]{[#1]}%
\providecommand \BibitemOpen [0]{}%
\providecommand \bibitemStop [0]{}%
\providecommand \bibitemNoStop [0]{.\EOS\space}%
\providecommand \EOS [0]{\spacefactor3000\relax}%
\providecommand \BibitemShut  [1]{\csname bibitem#1\endcsname}%
\let\auto@bib@innerbib\@empty
\bibitem [{\citenamefont {De}\ \emph {et~al.}(2018)\citenamefont {De},
  \citenamefont {Finstad}, \citenamefont {Lattimer}, \citenamefont {Brown},
  \citenamefont {Berger},\ and\ \citenamefont {Biwer}}]{De:2018}%
  \BibitemOpen
  \bibfield  {author} {\bibinfo {author} {\bibfnamefont {S.}~\bibnamefont
  {De}}, \bibinfo {author} {\bibfnamefont {D.}~\bibnamefont {Finstad}},
  \bibinfo {author} {\bibfnamefont {J.~M.}\ \bibnamefont {Lattimer}}, \bibinfo
  {author} {\bibfnamefont {D.~A.}\ \bibnamefont {Brown}}, \bibinfo {author}
  {\bibfnamefont {E.}~\bibnamefont {Berger}}, \ and\ \bibinfo {author}
  {\bibfnamefont {C.~M.}\ \bibnamefont {Biwer}},\ }\href {\doibase
  10.1103/PhysRevLett.121.091102} {\bibfield  {journal} {\bibinfo  {journal}
  {Phys. Rev. Lett.}\ }\textbf {\bibinfo {volume} {121}},\ \bibinfo {pages}
  {091102} (\bibinfo {year} {2018})}\BibitemShut {NoStop}%
\bibitem [{\citenamefont {Biwer}\ \emph {et~al.}(2018)\citenamefont {Biwer},
  \citenamefont {Capano}, \citenamefont {De}, \citenamefont {Cabero},
  \citenamefont {Brown}, \citenamefont {Nitz},\ and\ \citenamefont
  {Raymond}}]{Biwer:2018osg}%
  \BibitemOpen
  \bibfield  {author} {\bibinfo {author} {\bibfnamefont {C.~M.}\ \bibnamefont
  {Biwer}}, \bibinfo {author} {\bibfnamefont {C.~D.}\ \bibnamefont {Capano}},
  \bibinfo {author} {\bibfnamefont {S.}~\bibnamefont {De}}, \bibinfo {author}
  {\bibfnamefont {M.}~\bibnamefont {Cabero}}, \bibinfo {author} {\bibfnamefont
  {D.~A.}\ \bibnamefont {Brown}}, \bibinfo {author} {\bibfnamefont {A.~H.}\
  \bibnamefont {Nitz}}, \ and\ \bibinfo {author} {\bibfnamefont
  {V.}~\bibnamefont {Raymond}},\ }\href@noop {} {\  (\bibinfo {year} {2018})},\
  \Eprint {http://arxiv.org/abs/1807.10312} {arXiv:1807.10312 [astro-ph.IM]}
  \BibitemShut {NoStop}%
\bibitem [{\citenamefont {Foreman-Mackey}\ \emph {et~al.}(2013)\citenamefont
  {Foreman-Mackey}, \citenamefont {Hogg}, \citenamefont {Lang},\ and\
  \citenamefont {Goodman}}]{emcee}%
  \BibitemOpen
  \bibfield  {author} {\bibinfo {author} {\bibfnamefont {D.}~\bibnamefont
  {Foreman-Mackey}}, \bibinfo {author} {\bibfnamefont {D.~W.}\ \bibnamefont
  {Hogg}}, \bibinfo {author} {\bibfnamefont {D.}~\bibnamefont {Lang}}, \ and\
  \bibinfo {author} {\bibfnamefont {J.}~\bibnamefont {Goodman}},\ }\href
  {http://stacks.iop.org/1538-3873/125/i=925/a=306} {\bibfield  {journal}
  {\bibinfo  {journal} {Publications of the Astronomical Society of the
  Pacific}\ }\textbf {\bibinfo {volume} {125}},\ \bibinfo {pages} {306}
  (\bibinfo {year} {2013})}\BibitemShut {NoStop}%
\bibitem [{\citenamefont {Vousden}\ \emph {et~al.}(2016)\citenamefont
  {Vousden}, \citenamefont {Farr},\ and\ \citenamefont
  {Mandel}}]{vousden:2016}%
  \BibitemOpen
  \bibfield  {author} {\bibinfo {author} {\bibfnamefont {W.~D.}\ \bibnamefont
  {Vousden}}, \bibinfo {author} {\bibfnamefont {W.~M.}\ \bibnamefont {Farr}}, \
  and\ \bibinfo {author} {\bibfnamefont {I.}~\bibnamefont {Mandel}},\ }\href
  {\doibase 10.1093/mnras/stv2422} {\bibfield  {journal} {\bibinfo  {journal}
  {Monthly Notices of the Royal Astronomical Society}\ }\textbf {\bibinfo
  {volume} {455}},\ \bibinfo {pages} {1919} (\bibinfo {year}
  {2016})}\BibitemShut {NoStop}%
\bibitem [{\citenamefont {Goodman~J.}(2010)}]{mcmc}%
  \BibitemOpen
  \bibfield  {author} {\bibinfo {author} {\bibfnamefont {W.~J.}\ \bibnamefont
  {Goodman~J.}},\ }\href {\doibase 10.2140/camcos.2010.5.65} {\bibfield
  {journal} {\bibinfo  {journal} {Commun. Appl. Math. Comput. Sci.}\ }\textbf
  {\bibinfo {volume} {5}},\ \bibinfo {pages} {65} (\bibinfo {year}
  {2010})}\BibitemShut {NoStop}%
\bibitem [{\citenamefont {Liu}\ \emph {et~al.}(2016)\citenamefont {Liu},
  \citenamefont {Elshall}, \citenamefont {Ye}, \citenamefont {Beerli},
  \citenamefont {Zeng}, \citenamefont {Lu},\ and\ \citenamefont
  {Tao}}]{Liu:2016}%
  \BibitemOpen
  \bibfield  {author} {\bibinfo {author} {\bibfnamefont {P.}~\bibnamefont
  {Liu}}, \bibinfo {author} {\bibfnamefont {A.~S.}\ \bibnamefont {Elshall}},
  \bibinfo {author} {\bibfnamefont {M.}~\bibnamefont {Ye}}, \bibinfo {author}
  {\bibfnamefont {P.}~\bibnamefont {Beerli}}, \bibinfo {author} {\bibfnamefont
  {X.}~\bibnamefont {Zeng}}, \bibinfo {author} {\bibfnamefont {D.}~\bibnamefont
  {Lu}}, \ and\ \bibinfo {author} {\bibfnamefont {Y.}~\bibnamefont {Tao}},\
  }\href {\doibase 10.1002/2014WR016718} {\bibfield  {journal} {\bibinfo
  {journal} {Water Resources Research}\ }\textbf {\bibinfo {volume} {52}},\
  \bibinfo {pages} {734} (\bibinfo {year} {2016})}\BibitemShut {NoStop}%
\end{thebibliography}
